\documentclass[twocolumn]{jpsj3} 
\usepackage{bm,amssymb,amsmath}
\usepackage{graphicx}   
\usepackage{txfonts}
\usepackage{braket}
\usepackage[table]{xcolor}
\usepackage{arydshln}
\newcommand{\simg}{\stackrel{>}{_\sim}}
\newcommand{\siml}{\stackrel{<}{_\sim}}

\makeatletter
\newcommand{\figcaption}[1]{\def\@captype{figure}\caption{#1}}
\newcommand{\tblcaption}[1]{\def\@captype{table}\caption{#1}}
\makeatother

\newcommand\pb[1]{%
\begin{tabular}{@{}l@{}}#1\end{tabular}}

\title{
Derivation of RKKY Interaction between Multipole Moments in CeB$_6$ by the Effective Wannier Model based on the Bandstructure Calculation
}
\author{Takemi YAMADA\thanks{E-mail address: t-yamada@rs.tus.ac.jp}
and 
Katsurou HANZAWA
}
\inst{
Department of Physics, Faculty of Science and Technology, Tokyo University of Science, Chiba, 278-8510, Japan
}
\abst{
We have investigated the electronic states of CeB$_6$ and have directly calculated the RKKY interaction on the basis of the 74-orbital effective Wannier model which includes 14 Ce-$f$ orbitals and 60 conduction ($c$) orbitals of Ce-$d,s$ and B-$p,s$ derived from the density-functional theory bandstructure calculation. 
By using not only the $c$-band dispersion but also the $f$-$c$ mixing matrix elements of the Wannier model, 
the realistic couplings for all 15 active multipole moments in $\Gamma_8$ quartet subspace are obtained in the wavevector $\bm{q}$-space and real-space. Both of the $\Gamma_{5g}$ quadrupoles $(O_{yz},O_{zx},O_{xy})$ and the $\Gamma_{2u}$ octupole $T_{xyz}$ couplings are maximally enhanced with $\bm{q}=(\pi,\pi,\pi)$ which naturally explains the phase II of the antiferro-quadrupolar ordering at $T_{Q}=3.2$ K, 
and are also enhanced with $\bm{q}=(0,0,0)$ corresponding to the elastic softening of $C_{44}$. 
Also the couplings of the $\Gamma_{5u}$ octupoles $T_{x}^{\beta}$, $T_{y}^{\beta}$ and $T_{z}^{\beta}$ are quite large for 
$\bm{q}=(\pi,0,0)$, $(0,\pi,0)$ and $(0,0,\pi)$, which yields the antiferro-octupolar ordering of a possible candidate for phase IV of Ce$_{x}$La$_{1-x}$B$_6$. 
The intersite vector dependence of the RKKY couplings exhibit 
different long-range, oscillating, isotropic and anisotropic behaviors depending on the types of the multipole moments. 
The present approach enables us to provide the information about the possible multipole ordering 
in an unbiased way and is easily available for other localized $f$ electron materials 
once the $c$ states and $f$-$c$ mixing elements are given from the bandstructure calculation.
}

\begin{document}
\maketitle

\section{Introduction}\label{sec1}

\begin{figure*}[t]
\centering
\includegraphics[width=17.0cm]{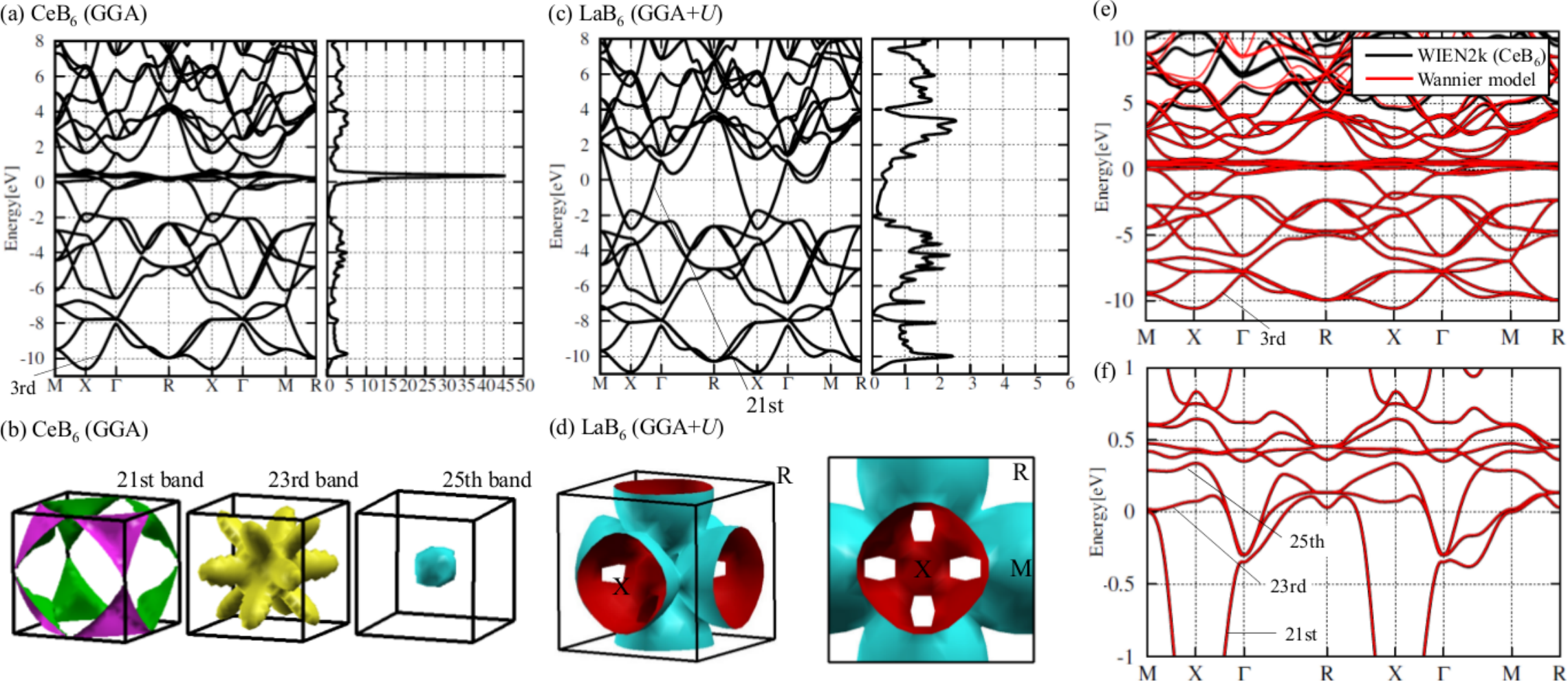}
\caption{(Color online) 
(a)-(d) DFT bandstructures with the DOS and FSs for CeB$_6$ [(a), (b)] and LaB$_6$ [(c), (d)] in the simple cubic BZ, 
where the high-symmetry points are $\Gamma$[$(0,0,0)$], X[$(\pi,0,0)$], M[$(\pi.\pi,0)$] and R[$(\pi.\pi,\pi)$]. 
(e), (f) The comparison between the 74-orbital effective Wannier model and DFT bands of CeB$_6$. 
}
\label{Fig01}
\end{figure*}

CeB$_6$ has been known as a typical and remarkable compound exhibiting a rich phase diagram 
of the multipole orderings
\cite{Santini2009,Kuramoto2009,Kusunose2008,Cameron2016} 
and extensively studied experimentally
\cite{
Cameron2016,
Fujita1980,
Kawakami1980,
Takase1980,
Zirngiebl1984,
Furuno1985,
Bredl1987,
Effantin1982,
Effantin1985,
Erkelens1987,
Takigawa1983,
Luthi1984,
Goto1985,
Nakamura1994,
Nakamura1995,
Tayama1997,
Hiroi1997,
Sakai1997,
Friemel2012,
Jang2014,
Jang2017,
Nikitin2018,
Onuki1989,
Endo2006,
Souma2004,
Neupane2015,
Ramankuttya2016,
Koitzsch2016
} 
and theoretically\cite{
Ohkawa1983,
Ohkawa1985,
Shiina1997,
Shiina1998,
Shiba1999,
Hanzawa2000,
Sakurai-Kuramoto2004,
Sakurai2005,
Lu-Huang2017,
Barman2019
}.
Due to the large spin-orbit coupling (SOC) and the cubic crystalline electric field (CEF), 
the ground state of 4$f^1$ in Ce$^{3+}$ ion is the $\Gamma_8$ quartet separated from the excited $\Gamma_7$ doublet by 540 K\cite{Zirngiebl1984}, and has a inherently the degrees of freedom of 15 active multipole moments as shown in Table \ref{table01}. 

\begin{table}[b]
\vspace{-0.4cm}
\centering
\scalebox{0.89}{
\begin{tabular}{cccc}
\hline \rule{0pt}{4mm}
IRR [dimension] & vector & pseudospin & multipole \\[2pt] \hline\hline \rule{0pt}{4mm}
$\Gamma_{2u}$ [1] & $\xi$ & $\tau^{y}$ & $\frac{2}{9\sqrt{5}}T_{xyz}$ \\[2pt] \rule{0pt}{4mm}
$\Gamma_{3g}$ [2] & $\bm{\tau}\rq{}$ & $(\tau^{z},\tau^{x})$ & $\frac{1}{4}(O_{u},O_{v})$ \\[2pt] \rule{0pt}{4mm}
$\Gamma_{5g}$ [3] & $\bm{\mu}$ & $(\tau^{y}\sigma^{x},\tau^{y}\sigma^{y},\tau^{y}\sigma^{z})$ & $(O_{yz},O_{zx},O_{xy})$ \\[2pt] \rule{0pt}{4mm}
$\Gamma_{4u}^{(1)}$ [3]& $\bm{\sigma}$ &$(\sigma^{x},\sigma^{y},\sigma^{z})$ &$\frac{14}{15}\bm{J}-\frac{4}{45}\bm{T}^{\alpha}$ \\[2pt] \rule{0pt}{4mm}
$\Gamma_{4u}^{(2)}$ [3]& $\bm{\eta}$ &$(\eta^{+}\sigma^{x},\eta^{-}\sigma^{y},\tau^{z}\sigma^{z})$ &$-\frac{2}{15}\bm{J}+\frac{7}{45}\bm{T}^{\alpha}$ \\[2pt] \rule{0pt}{4mm}
$\Gamma_{5u}$ [3] & $\bm{\zeta}$& $(\zeta^{+}\sigma^{x},\zeta^{-}\sigma^{y},\tau^{x}\sigma^{z})$ & $\frac{1}{3\sqrt{5}}\bm{T}^{\beta}$ \\[2pt]
\hline
\end{tabular}
}
\vspace{-0.3cm}
\caption{
The irreducible representations (IRRs) and notations for the active multipole moments in $\Gamma_8$  subspace\cite{Kusunose2008} 
where 
$\bm{J}~(\bm{T}^{\alpha,\beta})$ is the dipole (octupole), 
$\bm{J}=(J_{x},J_{y},J_{z})$, 
$\bm{T}^{\alpha(\beta)}=(T_{x}^{\alpha(\beta)},T_{y}^{\alpha(\beta)},T_{z}^{\alpha(\beta)})$, 
$\eta^{\pm}=-(\tau^{z}\mp\sqrt{3}\tau^{x})/2$, 
$\zeta^{\pm}=-\frac{1}{2}(\tau^{x}\pm\sqrt{3}\tau^{z})$, 
and $g~(u)$ means even (odd) time-reversal symmetry. 
In this paper we call $\frac{2}{9\sqrt{5}}T_{xyz}$, $\frac{1}{4}(O_{u},O_{v})$ and $\frac{1}{3\sqrt{5}}\bm{T}^{\beta}$ 
just as $T_{xyz}$, $(O_{u},O_{v})$ and $\bm{T}^{\beta}$, 
but all the multipole operators are normalized. 
}\label{table01}
\end{table} 

Up to now, three phases exist in temperature $T$ and external magnetic field $H$ plane of CeB$_6$. 
Normal phase (phase I) from a room temperature down to a few K with $H=0$ is a typical Kondo lattice metal with a highly-enhanced specific-heat coefficient\cite{Furuno1985,Bredl1987} $C/T=250$ mJ/mol$\cdot$K$^2$. 
With decreasing $T$, phase II emerges at a critical temperature $T_{Q}=3.2$ K with the ordering wavevector $\bm{q}=(\pi,\pi,\pi)$ 
and is confirmed by the antiferro-quadrupolar (AFQ) ordering of the $\Gamma_{5g}$ quadrupoles $(O_{yz},O_{zx},O_{xy})$. 
The ordering tendency of the $\Gamma_{5g}$ quadrupole moment is supported from the elastic-softening of $C_{44}$ at low temperature\cite{Luthi1984,Goto1985,Nakamura1994,Nakamura1995}. 
Interestingly, $T_{Q}$ increases with increasing the applied field $H$, 
where the $\Gamma_{2u}$ octupole $T_{xyz}$ moment is induced by $H$ in addition to the $\Gamma_{5g}$ quadrupoles, 
which is well understood by the analysis of NMR\cite{Sakai1997}. 
The phase III is a antiferro-magnetic (AFM) ordering of $\Gamma_{4u}$ magnetic moments $(\sigma^{x},\sigma^{y},\sigma^{z})$ at  $T_{N}=2.3$ K with the double-$\bm{q}$-structure of $\bm{Q}_{1}=(\tfrac{\pi}{2},\tfrac{\pi}{2},0)$ and $\bm{Q}_{2}=(\tfrac{\pi}{2},\tfrac{\pi}{2},\pi)$. 



4$f$ electron state of CeB$_6$ is believed to be almost localized in Ce$^{3+}$-ion from the several experiments of the magnetic and transport properties. 
More directly, the Fermi-surface (FS) has been observed in the de Haas-van Alphen (dHvA) experiments\cite{Onuki1989,Endo2006}, the angle resolved photoemission spectroscopy (ARPES)\cite{Souma2004,Neupane2015,Ramankuttya2016} 
and the high-resolution photoemission tomography\cite{Koitzsch2016}. 
They has indicated an ellipsoidal FS centered at X point in the Brillouin zone (BZ) 
which is almost the same as that of LaB$_6$ with the 4$f^{0}$ state. Hence the 4$f$ state in CeB$_6$ is localized and hardly participates in the formation of FS. 

In such a localized $f$ electron picture, 
Ruderman-Kittel-Kasuya-Yosida (RKKY) interaction\cite{RK1954,Kasuya1956,Yosida1957} 
plays an important role for the multipole ordering where the intersite coupling between the multipole moments of $f$ electrons is mediated by the itinerant $c$ band electrons\cite{SL1985,Frisken-Miller1986}. 
The RKKY model of CeB$_6$ was proposed by Ohkawa\cite{Ohkawa1983,Ohkawa1985} firstly, 
and later developed by Shiina {\it et al.,}\cite{Shiina1997,Shiina1998} 
where all 15 active multipole moments had been taken into account in correct symmetry, 
and reproduced the experimental $T$-$H$ phase diagram 
where only nearest neighbor couplings and the largest $\Gamma_{5g}$ quadrupoles couplings were assumed. 
This assumption was discussed from the symmetry of the RKKY couplings\cite{Shiba1999,Hanzawa2000}, 
but there was no explicit calculation for the signs and values of the couplings, and 
also no discussion about the long-range property of the RKKY multipole coupling of CeB$_6$. 
Later Sakurai {\it et al.,}\cite{Sakurai-Kuramoto2004,Sakurai2005} studied the RKKY multipole couplings of CeB$_6$ microscopically 
such as 
the effect of the $f^0$ and $f^2$ intermediate states, $c$ band number dependence 
and the ratio of the $f$-$c$ mixing elements described by the Slater-Koster (SK) parameters, 
but plausible ordering moment types and wavevectors could not be obtained. 

As is often discussed in the RKKY mechanism, 
the $c$ band states and their couplings with the $f$ states in the realistic materials must be important for determining the ordering moment types and wavevectors. 
Therefore the microscopic description of the $c$ band states and $f$-$c$ mixing elements from the realistic bandstructure calculation is needed, though such studies are quite limited\cite{Tanaka2010,Hanzawa2015}. 
In these studies\cite{Tanaka2010,Hanzawa2015}, the $c$ states is described by the Wannier orbitals obtained from the bandstructure calculation 
but the $f$-$c$ mixing elements using the calculation of the RKKY coupling are treated by the SK parameters only with the nearest neighbor sites, where several arbitrary parameters and assumptions are included. 
Hence more decisive and widely-applicable approach reflecting the individual material properties is highly desired.

In this paper, we study the electronic states of CeB$_6$ and calculate the RKKY interaction based on the 74-orbital effective Wannier model derived from the bandstructure calculation directly. 
In Sec. \ref{sec2}, we calculate the bandstructures of CeB$_6$ and LaB$_6$ and construct the effective Wannier model of CeB$_6$, and examine the quasi-particles states and their multipole fluctuations based on the renormalized Wannier model in Sec. \ref{sec3}. 
Next in Sec. \ref{sec4}, 
we formulate the present RKKY mechanism based on the multi-orbital Kondo lattice model with both of 
the $\Gamma_{8}$ quartet and 60 $c$ orbitals, 
and present the results of the RKKY multipole couplings for all moments as functions of the wavevector and intersite vectors. 
Finally we give the summary and discussion in Sec. \ref{sec5}.



\section{Bandstructure calculation \& Wannier model}\label{sec2}

\subsection{DFT Bandstructure calculation of CeB$_6$ \& LaB$_6$}
First we calculate the electronic states of CeB$_6$ and LaB$_6$ 
by using the WIEN2k code\cite{w2k2002,w2k-Schwarz1990,w2k-Schwarz2002
}, based on the framework of the density-functional theory (DFT) with 
the generalized gradient approximation (GGA)
\cite{PBE-GGA1996}. 
The SOC is fully included within the second variation approximation. The crystallographical parameters are the space group $Pm\bar{3}m$ (No. 221), the lattice constant $a=4.141{\rm \AA}$ and the internal coordinates $(x/a,y/b,z/c)=(0,0,0)$ for Ce and $(\frac{1}{2},\frac{1}{2},u)$ for B with $u=(\sqrt{2}-1)/2\sim 0.2071$
\cite{Wyckoff1964}. 
In self-consistent calculation, we use 156 $k$-points in the irreducible part of the simple cubic BZ, 
the muffin-tin radii $R_{\rm MT}=2.50~(1.62)$ a.u. for Ce (B) and the plane-wave cuttoff of $R_{\rm MT}K_{\rm max}=8$. 
For the calculation of LaB$_6$, we use the same parameters of CeB$_6$ but employ the GGA+$U$ method 
with $U=60$ eV 
for La-$f$ level 
so as to eliminate the $f$ weights in the $c$ bands, 
since we focus the pure $c$ band state of CeB$_6$ not bulk property of LaB$_6$. 

The obtained bandstructures with the density-of-states (DOSs) and FSs are shown in Fig.\ref{Fig01} for CeB$_6$ [(a) \& (b)] and LaB$_6$ [(c) \& (d)]. 
In CeB$_6$, the large $f$ contribution due to the 14 $f$ spin-orbital states around Fermi energy ($E_{\rm F}$) is observed with  the strong peak of DOS as shown in the right panel of Fig. \ref{Fig01} (a). 
On the other hands in LaB$_6$ 
the $f$ states is absent in the bandstructure and DOS [Fig. \ref{Fig01} (b)] as expected due to the effect of the GGA+$U$. 
Except for the $f$ band states, the global bandstructures of CeB$_6$ and LaB$_6$ are closely resembled below and above $E_{\rm F}$. 
The calculated FSs of CeB$_6$ and LaB$_6$ are plotted in Figs. \ref{Fig01} (b) and (d), respectively. 
Three FSs are obtained from the 21st, 23rd and 25th bands for CeB$_6$ 
while for LaB$_6$ an ellipsoidal FS centered at X point slightly connected each other is obtained from the 21th-band. 
Here we note that 
all bands have two-folded degeneracy due to the time-reversal symmetry 
and two additional bands (1st and 2nd bands) are located in $E_{\rm F}-15$ eV (not shown) 
which are the lowest bands in the Wannier model in next subsection. 

\subsection{Construction of Wannier model for CeB$_6$}
Next we construct the 74-orbital effective Wannier model based on the maximally localized Wannier functions (MLWFs) method\cite{MV1997,Souza2001,Marzari2012,w90-Mostofi2008,w2w-Kunes2010} from the DFT bandstructure of CeB$_6$, 
where we prepare 14 $f$-states from Ce-$f$ (7 orbital $\times$ 2 spin) 
and 60 $c$-states from Ce-$d$ (5 orbital $\times$ 2 spin), Ce-$s$ (1 orbital $\times$ 2 spin), B-$p$ (6 site $\times$ 3 orbital $\times$ 2 spin) and B-$s$ (6 site $\times$ 1 orbital $\times$ 2 spin) as basis functions, and 
set considerably wide energy window 
in order to ensure the good localization of Wannier orbitals in the disentanglement procedure. 
The obtained bandstructure of the Wannier model is 
plotted in Figs. \ref{Fig01} (e) and (f) together with the DFT bandstructure of CeB$_6$ (black), 
where the Wannier model is well reproduced the DFT bandstructure upto $E_{\rm F}+4$ eV 
and the shapes of the Wannier orbitals are similar to the atomic-orbitals significantly. 


The obtained model can be written by the following tight-binding (TB) Hamiltonian as, 
\begin{align}
H_{\rm TB}&=\sum_{ij}\sum_{mm\rq{}}h_{im,jm\rq{}}^{ff}f_{im}^{\dagger}f_{jm\rq{}}^{} +\sum_{ij}\sum_{\ell\ell\rq{}}h_{i\ell,j\ell\rq{}}^{cc}c_{i\ell}^{\dagger}c_{j\ell\rq{}}^{}\nonumber\\
&+\sum_{ij}\sum_{m\ell}\left(V_{im,j\ell}f_{im}^{\dagger}c_{j\ell}^{}+h.c.\right),
\label{eq:HTB}
\end{align}
where $f_{im}^{\dagger}~(c_{i\ell}^{\dagger})$ is a creation operator for a $f~(c)$ electron with unit-cell $i$
and 14 (60) spin-orbital states $m~(\ell)$. 
Here 14 $f$ states of $m$ are represented by the CEF eigenstates as $\Gamma_{8}$ quartet and $\Gamma_{7}$ doublet with the total angular momentum $J=5/2$, and $\Gamma_{6}$, $\Gamma_{7}$ doublets and $\Gamma_{8}$ quartet with $J=7/2$. 
The $f$-$f$ ($c$-$c$) matrix element of $h_{im,jm\rq{}}^{ff}~(h_{i\ell,j\ell\rq{}}^{cc})$ 
includes the $f~(c)$ energy levels, SOC couplings, CEF splittings and $f$-$f$ ($c$-$c$) hopping integrals, 
and $V_{im,j\ell}$ is the $f$-$c$ mixing element which is finite only for the intersite terms due to the inversion symmetry. 
The wavevector $\bm{k}$-representation of $H_{\rm TB}$ is given by, 
\begin{align}
H_{\rm TB}&=\sum_{\bm{k}}\sum_{mm\rq{}}h_{mm\rq{}}^{ff}(\bm{k})f_{\bm{k}m}^{\dagger}f_{\bm{k}m\rq{}}^{} +\sum_{\bm{k}}\sum_{\ell\ell\rq{}}h_{\ell\ell\rq{}}^{cc}(\bm{k})c_{\bm{k}\ell}^{\dagger}c_{\bm{k}\ell\rq{}}^{}\nonumber\\
&+\sum_{\bm{k}}\sum_{m\ell}\left(V_{\bm{k}m\ell}f_{\bm{k}m}^{\dagger}c_{\bm{k}\ell}^{}+h.c.\right)
=\sum_{\bm{k}s}\varepsilon_{\bm{k}s}a_{\bm{k}s}^{\dagger}a_{\bm{k}s}^{},
\label{eq:HTB-k}
\end{align}
where $\varepsilon_{\bm{k}s}$ is the eigenenergy with $\bm{k}$ and band-index $s$ 
and $a_{\bm{k}s}^{\dagger}$ is a creation operator for a electron with $\bm{k},s$, 
which is transformed into $m$ and $\ell$ states as $\displaystyle a_{\bm{k}s}=\sum_{m}u_{\bm{k}sm}f_{\bm{k}m}+\sum_{\ell}u_{\bm{k}s\ell}c_{\bm{k}\ell}$ 
where $u_{\bm{k}sm}~(u_{\bm{k}s\ell})$ is the eigenvector component of $m~(\ell)$ state. 


Several atomic parameters are obtained from the Wannier model, 
such as the SOC splitting for Ce-$4f$ between $J=5/2$ and $J=7/2$ states $\Delta_{\rm SOC}=0.33$ eV close to the experimental value of 3000 K, 
the atomic CEF splitting between $\Gamma_8$ and $\Gamma_7$, $\Delta_{\rm CEF}=8.2$ meV
which is smaller than the experimental value of 540 K (=$46$ meV). 
The $f~(c)$ electron number per unit-cell is $n^{f}=1.24$ ($n^{c}=20.86$) and the total number is $n_{tot}=22$. 
All the $f$ electron number for each CEF state becomes finite 
where $n^{f}(\Gamma_8)=0.634$ and $n^{f}(\Gamma_7)=0.205$ for $J=5/2$
and $n^{f}(\Gamma_6)=0.098$, $n^{f}(\Gamma_7)=0.088$, and $n^{f}(\Gamma_8)=0.216$ for $J=7/2$, 
due to the considerable $f$-$f$ hopping and $f$-$c$ mixing, 
which is indispensable within the DFT-based calculation.

\section{Quasi-particle band states \& Multipole fluctuations}\label{sec3}
\subsection{Renormalized tight-binding model}
As mentioned in Sec. \ref{sec2}, the $f$ electron state obtained here is fully itinerant 
and differs from the expected situation in the real material as $n^{f}(\Gamma_8)\sim 1$. 
In this section, we examine the change of the electronic states and its multipole fluctuations 
from the itinerant $f$ band state to the localized $f$ state when $n^{f}(\Gamma_8)=1$ in the realistic CeB$_6$ bandstructures. 
For this purpose, we introduce a renormalization factor $Z_{m}^{f}$, which is explicitly derived from the Fermi-liquid (FL) theory
\cite{Yosida-Yamada1986}, 
where the many-body correlation effect of the local $f$-$f$ Coulomb interaction 
is introduced through the self-energy $\Sigma_{m}^{f}(\bm{k},\varepsilon)$ 
which is almost local $\Sigma_{m}^{f}(\bm{k},\varepsilon)=\Sigma_{m}^{f}(\varepsilon)$ 
and can be expanded around $\varepsilon=0$ by the following form, 
\begin{align}
&\Sigma_{m}^{f}(\varepsilon)=\Delta\varepsilon_{m}^{f}+\left(1-\frac{1}{Z_{m}^{f}}\right)\varepsilon - i\gamma_{m}\varepsilon^2 +O(\varepsilon^3),\\
&\Delta\varepsilon_{m}^{f}={\rm Re}\Sigma_{m}^{f}(0),~~Z_{m}^{f}=\left(1-\frac{d}{d\varepsilon}{\rm Re}\Sigma_{m}^{f}(\varepsilon)\Bigr|_{\varepsilon=0}\right)^{-1},
\end{align}
where $Z_{m}^{f}$ corresponds to an inverse mass-enhancement $m/m^{*}$, 
and $\Delta\varepsilon_{m}^{f}$ and $\gamma_{m}$ are a shift of the $f$ energy-level and a dumping rate of the quasi-particles respectively. 
Hence in the itinerant quasi-particle picture, our original model of $H_{\rm TB}$ is renormalized by $Z_{m}^{f}$ and $\Delta\varepsilon_{m}$, yielding the renormalized tight-binding model $H_{\rm RTB}$ as explicitly given by, 
\begin{align}
H_{\rm RTB}&=\sum_{ij}\sum_{mm\rq{}}\tilde{h}_{im,jm\rq{}}^{ff}f_{im}^{\dagger}f_{jm\rq{}}^{} +\sum_{ij}\sum_{\ell\ell\rq{}}h_{i\ell,j\ell\rq{}}^{cc}c_{i\ell}^{\dagger}c_{j\ell\rq{}}^{}\nonumber\\
&+\sum_{ij}\sum_{m\ell}\left(\tilde{V}_{im,j\ell}f_{im}^{\dagger}c_{j\ell}^{}+h.c.\right)
\label{eq:H-RTB}
\end{align}
where the renormalized $f$-$f$ ($f$-$c$) matrix elements $\tilde{h}_{im,jm\rq{}}^{ff}~(\tilde{V}_{im,j\ell})$ are written as,
\begin{align}
&\tilde{h}_{im,jm\rq{}}^{ff}=\left\{
\begin{array}{cc}
\varepsilon_{m}^{f}+\Delta\varepsilon_{m}^{f} & i=j,~m=m\rq{} \\
\sqrt{Z_{m}^{f}Z_{m\rq{}}^{f}}h_{im,jm\rq{}}^{ff} & i\neq j \\
\end{array}\right.\label{eq:hff}\\
&\tilde{V}_{im,j\ell}=\sqrt{Z_{m}^{f}}V_{im,j\ell}\label{eq:hfc}
\end{align}
where $\varepsilon_{m}^{f}$ is a $f$ energy-level of the CEF state $m$, where $h^{ff}_{im,im\rq{}}=0$ for $m\neq m\rq{}$,
and the $m$-dependence of $Z_{m}^{f}$ and $\Delta\varepsilon_{m}^{f}$ are dropped for simplicity as $Z_{m}^{f}=Z_{f}$ and  $\Delta\varepsilon_{m}^{f}=\Delta\varepsilon_{f}$, 
where $\Delta\varepsilon_{f}$ is set to $\Delta\varepsilon^{f}=0.27$ eV
so as to satisfy $n^{f}(\Gamma_8)=1$ and $n^{c}=21$ at $Z_{f}=0$. 
Hence the $f$-$f$ ($f$-$c$) hopping elements 
are renormalized by $Z_{f}~(\sqrt{Z_f})$. 
Throughout the calculation, we determine a chemical potential $\mu$ so as to satisfy 
$n_{tot}=n^{f}+n^{c}=22$ 
with 64$^3$ $\bm{k}$-meshes in the entire BZ.

\begin{figure}[t]
\centering
\includegraphics[width=8.7cm]{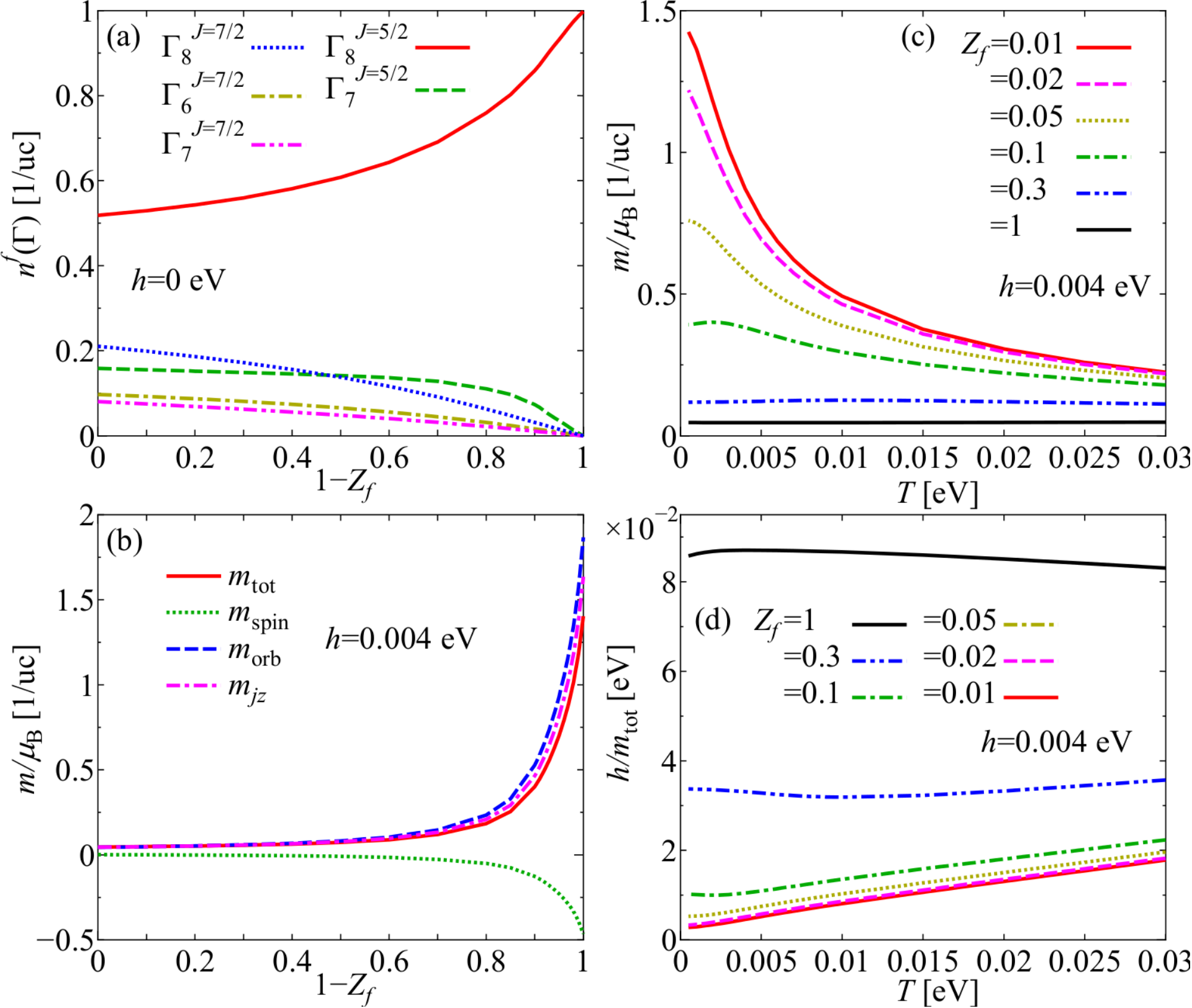}
\caption{(Color online) 
(a), (b): $Z_{f}$-dependence of (a) the $f$ electron number of each $f$ CEF state when $h=0$ eV; and  
(b) the $f$ electron magnetizations of total, spin, orbital and total angular momentum of $J=5/2$, $m_{\rm tot}$, $m_{\rm spin}$, $m_{\rm orb}$ and $m_{jz}$ with finite magnetic field $h=0.004$ eV. 
}
\label{Fig02}
\end{figure}

\subsection{Renormalized electronic states}
Figure \ref{Fig02} (a) shows the $Z_{f}$-dependence of $f$ electron number per CEF eigenstates
$n^{f}(\Gamma)$ with $T=0.002$ eV, 
where $\Gamma=\Gamma_{8}^{J=5/2}$, $\Gamma_{7}^{J=5/2}$ $\Gamma_{6}^{J=7/2}$, $\Gamma_{7}^{J=7/2}$
and $\Gamma_{8}^{J=7/2}$, and 
$Z_{f}=1$ ($Z_{f}=0$) corresponds to the DFT-band (localized $f$) limit. 
With decreasing $Z_{f}$, $n^{f}(\Gamma_{8}^{J=5/2})$ increases and finally becomes $n^{f}(\Gamma_{8}^{J=5/2})=1$ when $Z_{f}=0$ 
while $n^{f}(\Gamma_{7}^{J=5/2})$ and all other $n^{f}(\Gamma^{J=7/2})$ 
decrease and reach zero at $Z_{f}=0$. 
The change of $n^{f}(\Gamma)$ is rapidly for $Z_{f}\siml 0.1$ where the effective mass-enhancement reaches $m^{*}/m\simg 10$. 
The $f$ electron magnetization $\displaystyle m_{\rm tot}=m_{\rm spin}+m_{\rm orb}$ as a function of $Z_{f}$ is also plotted 
in Fig. \ref{Fig02} (b) together with its spin, orbital and $J_{z}$-components $m_{\rm spin}$, $m_{\rm orb}$ and $m_{jz}$, respectively, 
where the magnetic field is applied along the $z$-direction with $h=\mu_{\rm B}H=0.004$ eV. 
The Zeemann Hamiltonian is given by $H_{Z}=\left(\sigma^{z}+\ell^{z}\right)h$, and $m_{\rm spin}$ and $m_{\rm orb}$ are explicitly written as, 
\begin{align}
&m_{\rm spin}
=-\mu_{\rm B}\frac{1}{N}\sum_{\bm{k}s}\sum_{mm\rq{}}\left(\sigma^{z}\right)_{mm\rq{}}u_{\bm{k}sm}u_{\bm{k}sm\rq{}}^{*}f\left(\varepsilon_{\bm{k}s}\right),\\
&m_{\rm orb}
=-\mu_{\rm B}\frac{1}{N}\sum_{\bm{k}s}\sum_{mm\rq{}}\left(\ell^{z}\right)_{mm\rq{}}u_{\bm{k}sm}u_{\bm{k}sm\rq{}}^{*}f\left(\varepsilon_{\bm{k}s}\right),
\end{align}
where $\sigma^{z}~(\ell^{z})$ is a $z$-component of spin Pauli (orbital angular momentum) matrix for $m$-basis 
and $f(x)$ is the Fermi distribution function $\displaystyle f(x)=\frac{1}{e^{\beta (x-\mu)}+1}$. 
With decreasing $Z_{f}$, $m_{\rm tot}$ increases and finally reaches the saturated value of the $\Gamma_8$ state as 1.5$\mu_{\rm B}$ together with an opposite sign between $m_{\rm orb}$ and $m_{\rm spin}$ due to the SOC effect.

The $T$-dependence of the magnetization $m_{\rm tot}$ and inverse magnetization $h/m_{\rm tot}$ 
for several values of $Z_{f}$ are plotted in Figs. \ref{Fig02} (c) and (d) respectively. 
For $Z_{f}=1\sim 0.3$ the weak $T$-dependence of $m_{\rm tot}$ is observed as a Pauli paramagnetic behavior of the itinerant $f$ electron, while for $Z_{f}\siml 0.1$ $m_{\rm tot}$ increases with decreasing $T$, exhibiting the Curie paramagnetic behavior of the localized $f$ electron $m_{\rm tot}/h\sim 1/T$, 
which is more clearly observed in the inverse magnetization $h/m_{\rm tot}$ with a linear $T$-dependence. 
In such situations for $Z_{f}=0.1\sim0.01$, 
the electronic state is similar to the purely localized $f$ electron state on a single Ce-ion usually analyzed in the experiments. 
However in this study the $f$-$c$ mixings are still finite and the quasi-particle hybridization bands are formed with the wide-bandwidth $c$ band dispersion having the ellipsoidal FS observed ARPES of CeB$_6$.

\begin{figure}[t]
\centering
\includegraphics[width=8.7cm]{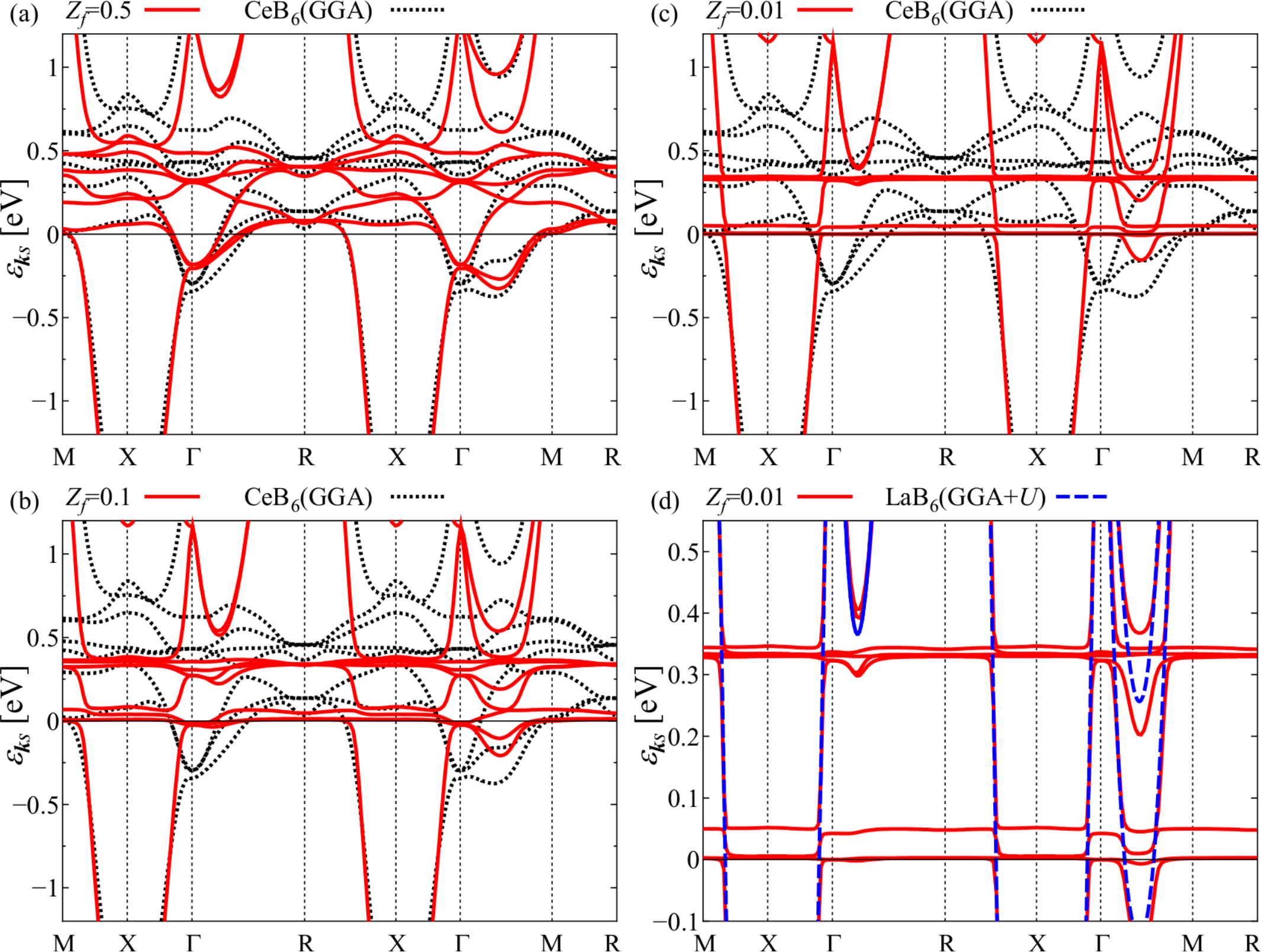}
\caption{(Color online) 
The change of the renormalized bandstructures (red solid line) for (a) $Z^{f}=0.5$ , (b) $Z^{f}=0.1$ and (c) $Z_{f}=0.01$ 
together with the DFT bandstructure of CeB$_6$ (black dotted line). 
(d) A comparison between the renormalized band $Z^{f}=0.01$ (red solid line) and the $f$-excluded band of LaB$_6$ by the GGA+$U$ (blue broken line). 
}
\label{Fig03}
\end{figure}

Next we check such renormalized bandstructures for several values of $Z_{f}$ as shown in Figs. \ref{Fig03} (a)-(d) 
together with the DFT-bandstructure of CeB$_6$ [Figs. \ref{Fig03} (a)-(c)] and 
the LaB$_6$ GGA+$U$ band without $f$ weights [Fig. \ref{Fig03} (d)]. 
From $Z_{f}=0.5$ [Fig. \ref{Fig03} (a)] to $Z_{f}=0.1$ [Fig. \ref{Fig03} (b)], 
the whole bandstructures are still close to the DFT-band of CeB$_6$ but their $f$ band-widths become narrow gradually,  exhibiting a separation between the lower $J=5/2$ bands and higher $J=7/2$ bands. 
In Fig. \ref{Fig03} (d) with $Z_{f}=0.01$ corresponding to $m^{*}/m\sim 100$, 
the almost flatted $J=7/2$ bands, and $\Gamma_7$ and $\Gamma_8$ bands of $J=5/2$ are clearly observed around $E_{\rm F}$, 
and they slightly hybridize with the wide-bandwidth $c$ bands expanding from the X point in the BZ.  
Interestingly, the $c$ band dispersion with $Z_{f}=0.01$ (red) is almost overlapping the LaB$_6$ band with the GGA+$U$ (black) as shown in Fig. \ref{Fig03} (d) except for the highly-flatted $f$ bands, resulting in the formation of almost the same FS of LaB$_6$. 
Hence the $c$ bands of CeB$_6$ with almost localized $f$ electron state
coincides that of LaB$_6$ without the La-$f$ contribution,  
and then their FS is also almost the same as that of LaB$_6$ as shown in Fig. \ref{Fig01} (d). 
These results strongly support the localized $f$ electron picture for CeB$_6$, and 
then the approach based on the periodic Anderson model and its perturbation w. r. t. the $f$-$c$ mixing 
is expected to giving a good starting point for treating this system.

\subsection{Multipole fluctuations in the quasi-particle bands}
Before going to the calculation of the RKKY interaction, we examine the multipole fluctuations under the renormalized $f$ bands on CeB$_6$ by calculating 
the multipole susceptibility $\chi_{O_{\Gamma}}^{}(\bm{q})$ with 
the multipole operator $O_{\Gamma}$ shown in Table \ref{table01} and the wavevector $\bm{q}$ which is given by, 
\begin{align}
&\chi_{O_{_{\Gamma}}}^{}(\bm{q})=\sum_{m_1m_2}\sum_{m_3m_4}O_{m_1m_2}^{\Gamma}O_{m_4m_3}^{\Gamma}\chi_{m_1m_2m_3m_4}^{}(\bm{q}),\\
&\chi_{m_1m_2m_3m_4}^{}(\bm{q})
=\frac{1}{N}\sum_{\bm{k}ss\rq{}}
u_{\bm{k}sm_3}^{*}
u_{\bm{k}sm_1}
u_{\bm{k}+\bm{q}s\rq{}m_2}^{*}
u_{\bm{k}+\bm{q}s\rq{}m_4}\nonumber\\
&\qquad\qquad\qquad
\times
\frac{
f(\varepsilon_{\bm{k}+\bm{q}s\rq{}})-f(\varepsilon_{\bm{k}s})}{
\varepsilon_{\bm{k}s}-\varepsilon_{\bm{k}+\bm{q}s\rq{}}},
\label{eq:chi-ff}
\end{align}
where $O_{\Gamma}=\sum_{mm\rq{}}O_{mm\rq{}}^{\Gamma}f_{im}^{\dagger}f_{im\rq{}}^{}$ and 
$O_{mm\rq{}}^{\Gamma}$ is the normalized 4$\times$4 matrix element of $O_{\Gamma}$ in $\Gamma_8$ subspace\cite{Kusunose2008}, and 
$\chi_{m_1m_2m_3m_4}^{}(\bm{q})$ is the irreducible $f$ electron susceptibility which depends on the distribution of $f$ states in the bandstructures through the renormalized $f$-$c$ mixing $\sqrt{Z_{f}}V_{im,j\ell}$. 

\begin{figure}[t]
\centering
\includegraphics[width=7.0cm]{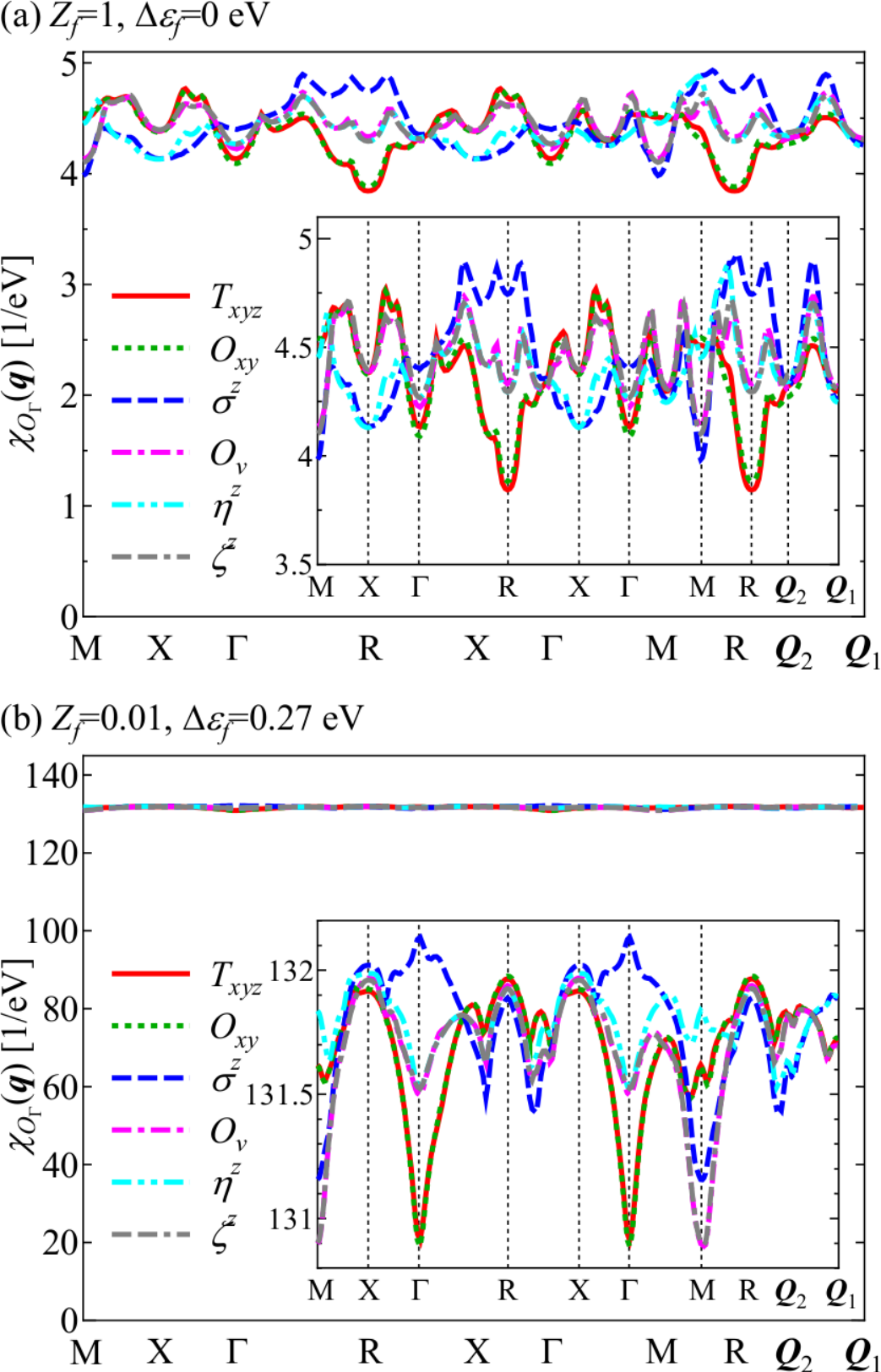}
\caption{(Color online) 
The multipole susceptibilities $\chi_{O_{\Gamma}}^{}(\bm{q})$ as a function of $\bm{q}$ along with high symmetry line in the BZ 
with (a) $Z_{f}=1$ \& $\Delta\varepsilon_{f}=0$ eV and (b) $Z_{f}=0.01$ \& $\Delta\varepsilon_{f}=0.27$ eV 
for several moments of 
$\Gamma_{2u}$ octupole $T_{xyz}$, 
$\Gamma_{5g}$ quadrupole $O_{xy}$, 
$\Gamma_{4u}^{(1)}$ multipole $\sigma^{z}$ corresponding to the magnetic moment, 
$\Gamma_{3g}$ quadrupole $O_{v}$, 
$\Gamma_{4u}^{(2)}$ multipole $\eta^{z}$, and
$\Gamma_{5u}$ octupole $\zeta^{z}$, 
where $\bm{Q}_{1}=(\tfrac{\pi}{2},\tfrac{\pi}{2},0)$ and $\bm{Q}_{2}=(\tfrac{\pi}{2},\tfrac{\pi}{2},\pi)$ are the AFM ordering vectors of phase III. The insets are their enlarged view. 
}
\label{Fig04}
\end{figure}

Figure \ref{Fig04} (a) shows the $\bm{q}$-dependence of $\chi_{O_{_{\Gamma}}}^{}(\bm{q})$ for each multipole moment 
with $Z_{f}=1$ and $\Delta\varepsilon_f=0$ eV corresponding to the DFT band limit as shown in Figs. \ref{Fig01} (e) and (f). 
The obtained $\bm{q}$-dependence is considerably weak and the explicit values of $\chi_{O_{_{\Gamma}}}^{}(\bm{q})$ fall within the only small range $\chi_{O_{_{\Gamma}}}^{}(\bm{q})=4\sim 5$ eV$^{-1}$ for all multipole moments and wavevectors $\bm{q}$. 
Among them the $\Gamma_{4u}^{(1)}$ magnetic multipole $\sigma^{z}$ susceptibility, 
where $\sigma^{x}$ and $\sigma^{y}$ are degenerate with $\sigma^{z}$, 
is barely large for the incommensurate wavevector around $\bm{q}=(\pi,\pi,\pi)$, 
while the $\Gamma_{5g}$ quadrupole $O_{xy}$ and the $\Gamma_{2u}$ ocutupole $T_{xyz}$ susceptibilities does not become large for the AFQ wavevector $\bm{q}=(\pi,\pi,\pi)$. 
The weak-$\bm{q}$ dependence of $\chi_{O_{_{\Gamma}}}^{}(\bm{q})$ 
becomes more notable for the almost localized $f$ case with $Z_{f}=0.01$ and  $\Delta\varepsilon_f=0.27$ eV as shown in Fig. \ref{Fig04} (b), 
where $\sigma^{z}$ becomes also maximum but its wavevector shifts to $\bm{q}=(0,0,0)$ 
as shown in the inset of Fig. \ref{Fig04} (b). 

In such a situation, the actual value of $\chi_{O_{_{\Gamma}}}^{}(\bm{q})$ becomes huge, 
where the extremely narrow $\Gamma_8$ bands are located in the very near and just above $E_{\rm F}$ with tiny $f$-$c$ mixing, and then the hybridized band $\varepsilon_{\bm{k}s}$ is highly degenerate for wide-range of the BZ, giving rise to the sizable enhancement of the Lindhard function of in Eq. (\ref{eq:chi-ff}). 
As far as such $\bm{q}$-independent $\chi_{O_{_{\Gamma}}}^{}(\bm{q})$, 
it is difficult to describe the development of the $(\pi,\pi,\pi)$-AFQ mode with $(O_{yz},O_{zx},O_{xy})$ 
by the perturbation of the $f$-$f$ Coulomb interaction such as the random phase approximation (RPA) and its extensions.

\section{RKKY Interaction of CeB$_6$}\label{sec4}

\begin{figure}[t]
\centering
\includegraphics[width=8.7cm]{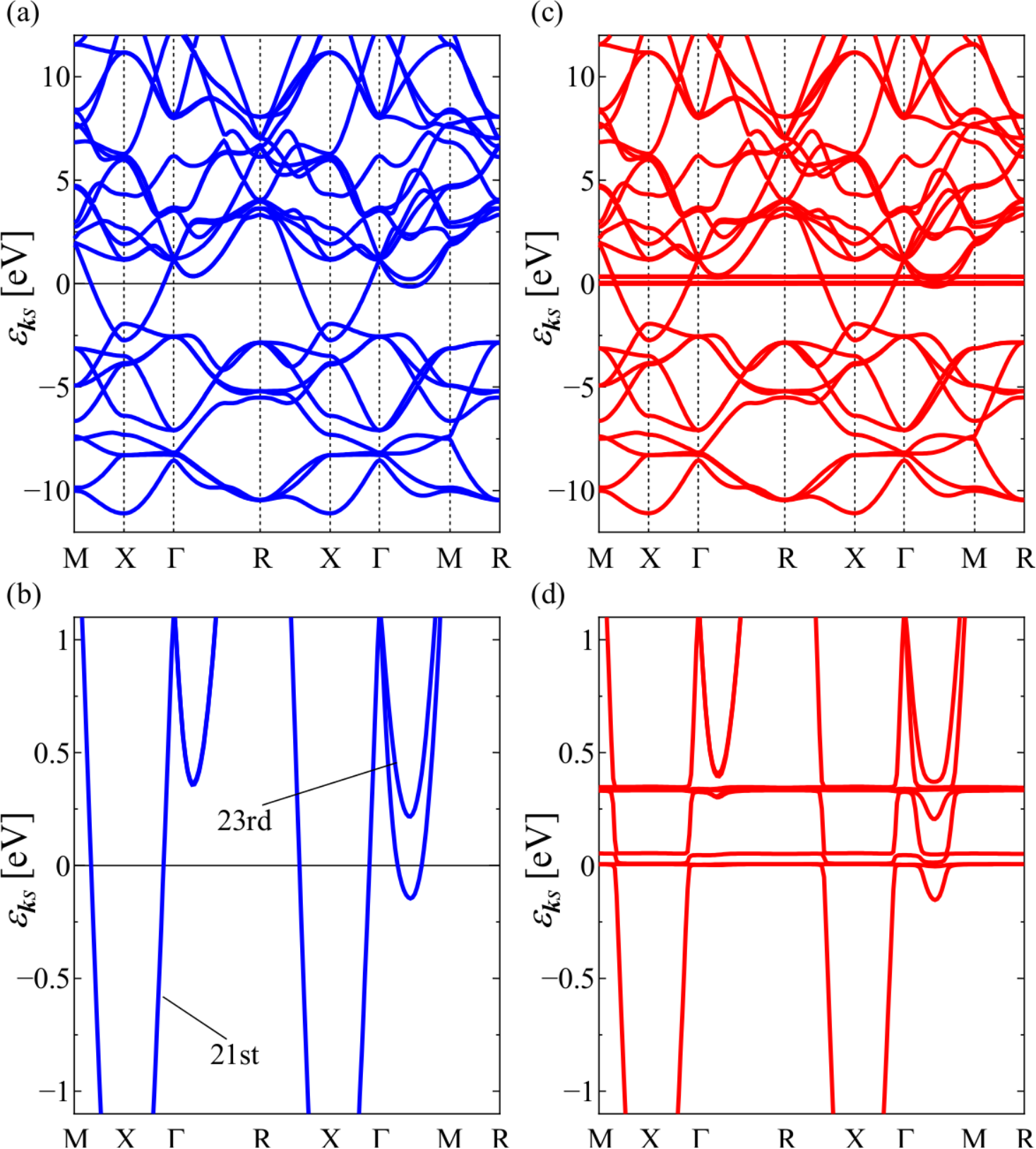}
\caption{(Color online) 
(a),(b): The $c$ bandstructure of the effective Wannier model for the RKKY coupling calculation without $f$ components.
(c),(d): The renormalized quasi-particles bandstructure for $Z_{f}=0.01$ with $f$ components.
}
\label{Fig05}
\end{figure}

\subsection{Derivation of RKKY Hamiltonian}
Here we consider the RKKY interaction between the multipole moments of $\Gamma_8$ quartet. 
For this purpose, we eliminate the $f$ energy-levels but use the $f$-$c$ mixing of the original Wannier model. 
The $c$ bandstructure for the calculation of RKKY couplings is shown in Figs. \ref{Fig05} (a) and (b), 
which is almost the same as that of LaB$_6$ as mentioned in Sec. \ref{sec3} 
and is compared to the strongly renormalized quasi-particles case with $Z_{f}=0.01$ as shown in Figs. \ref{Fig05} (c) and (d). 
During the calculation, $\mu$ is determined so as to keep $n_{tot}=n^{c}=21$ and $T$ is set to $T=0.005$ eV. 

The multi-orbital Kondo lattice Hamiltonian for the present model is given by, 
\begin{align}
H_{\rm MKL}&=
\sum_{im}\tilde{\varepsilon}_{m}^{f}f_{im}^{\dagger}f_{im}^{}
+\sum_{\bm{k}\ell\ell\rq{}}h_{\ell\ell\rq{}}^{cc}(\bm{k})c_{\bm{k}\ell}^{\dagger}c_{\bm{k}\ell\rq{}}^{}\nonumber\\
&+\sum_{i}\sum_{mm\rq{}}\sum_{\bm{k}\bm{k\rq{}}}\sum_{\ell\ell\rq{}}J_{imm\rq{}}^{\bm{k}\ell,
\bm{k}\rq{}\ell\rq{}}f_{im}^{\dagger}f_{im\rq{}}^{}c_{\bm{k}\ell}^{\dagger}c_{\bm{k}\rq{}\ell\rq{}}^{},
\end{align}
where $m$ represents 4-states in $\Gamma_8$ quartet $\Ket{m}=\Ket{1}\sim \Ket{4}$ 
with an degenerate energy-level $\tilde{\varepsilon}_{m}^{f}$, 
which are given with the $J_{z}$-base of $J=5/2$ $\Ket{JM}$ explicitly as,
$\Ket{1}=-\sqrt{\frac{1}{6}}\Ket{+\frac{3}{2}}-\sqrt{\frac{5}{6}}\Ket{-\frac{5}{2}}$, 
$\Ket{2}=\Ket{+\frac{1}{2}}$, 
$\Ket{3}=-\Ket{-\frac{1}{2}}$ and 
$\Ket{4}=\sqrt{\frac{1}{6}}\Ket{-\frac{3}{2}}+\sqrt{\frac{5}{6}}\Ket{+\frac{5}{2}}$. 
Here we note that 
$\tilde{\varepsilon}_{m}^{f}=\varepsilon_{m}^{f}+\Delta\varepsilon_{m}^{f}$ in Eq. (\ref{eq:hff}) for $m=1\sim4$ of $\Gamma_{8}$ 
and is pushed up from the bare $f$ energy-level $\varepsilon_{\Gamma_8}^{f}$ due to 
the DFT Hartree and GGA potentials which is of the order of a few eV. 
The $c$-$c$ matrix element $h_{\ell\ell\rq{}}^{cc}(\bm{k})$ 
includes the $c$ orbital energy $\varepsilon_{\bm{k}\ell}^{c}$ for $\ell=\ell\rq{}$ and the $c$-$c$ hopping $t_{\ell\ell\rq{}}^{cc}(\bm{k})$ for $\ell\neq\ell\rq{}$. 
The second term is rewritten by the $c$ band eigenstate 
$\displaystyle c_{\bm{k}s}=\sum_{\ell}u_{\bm{k}s\ell}^{c}c_{\bm{k}\ell}$ with the eigenenergy $\varepsilon_{\bm{k}s}$ and eigenvector $u_{\bm{k}s\ell}^{c}$. 

The Kondo coupling $J_{imm\rq{}}^{\bm{k}\ell,\bm{k}\rq{}\ell\rq{}}$ in the third term consists of the $f^{0}$- and $f^{2}$-intermediate process. 
In this paper, we take simple two assumptions for $J_{imm\rq{}}^{\bm{k}\ell,\bm{k}\rq{}\ell\rq{}}$; 
(1) only $f^0$-process is considered and the contribution of $f^2$-process is same as that of $f^0$-process 
and, 
(2) the scattered $c$ orbital energies are fixed to $\mu$, namely  $\varepsilon_{\bm{k}\ell}^{c},~\varepsilon_{\bm{k}\rq{}\ell\rq{}}^{c}\longrightarrow\mu$. Then the Kondo coupling $J_{imm\rq{}}^{\bm{k}\ell,\bm{k}\rq{}\ell\rq{}}$ can be written by the following simple form, 
\begin{align}
&J_{imm\rq{}}^{\bm{k}\ell,\bm{k}\rq{}\ell\rq{}}
=\frac{2}{N}
\frac{V_{\bm{k}m\ell}V_{\bm{k}\rq{}m\rq{}\ell\rq{}}^{*}}{\mu-\varepsilon_{\Gamma_8}^{f}}
e^{-i(\bm{k}-\bm{k}\rq{})\cdot\bm{R}_i}
\end{align}
where the prefactor 2 comes from the assumption (1) and 
$V_{\bm{k}m\ell}$ is the $\bm{k}$-represented $f$-$c$ mixing element in Eq. (\ref{eq:HTB-k}). 


The RKKY Hamiltonian can be obtained from the second-order perturbation w. r. t. the third term of $H_{\rm MKL}$ together with the thermal average for the $c$ states. The final form is given by,
\begin{align}
&H_{\rm RKKY}\!=\!-\!\sum_{\Braket{ij}}\!\sum_{m_1m_2}\!\sum_{m_3m_4}\!
K_{m_1m_2m_3m_4}(\bm{R}_{ij})
f_{im_1}^{\dagger}
f_{im_2}^{}
f_{jm_4}^{\dagger}
f_{jm_3}^{}
\label{eq:HRKKY-r},\\
&K_{m_1m_2m_3m_4}(\bm{R}_{ij})=\frac{1}{N}\sum_{\bm{q}}K_{m_1m_2m_3m_4}(\bm{q})~e^{i\bm{q}\cdot(\bm{R}_i-\bm{R}_j)},\label{eq:Kij}
\end{align}
where $K_{m_1m_2m_3m_4}(\bm{R}_{ij})$ is the RKKY coupling between the states $\{m_1,m_2\}$ at the unit-cell $\bm{R}_i$ and the states $\{m_3,m_4\}$ at $\bm{R}_j$ and $\Braket{ij}$ represents a summation for the intercell vectors $\bm{R}_{ij}=\bm{R}_i-\bm{R}_j$. 
The key quantity $K_{m_1m_2m_3m_4}(\bm{q})$ is given by,
\begin{align}
&K_{m_1m_2m_3m_4}(\bm{q})=\frac{1}{N}\sum_{\bm{k}ss\rq{}}\sum_{\ell_1\ell_2}\sum_{\ell_3\ell_4}
\frac{
V_{\bm{k}m_3\ell_3}^{*}
V_{\bm{k}m_1\ell_1}
V_{\bm{k}+\bm{q}m_2\ell_2}^{*}
V_{\bm{k}+\bm{q}m_4\ell_4}
}{(\mu-\varepsilon_{\Gamma_8}^{f})^2}\nonumber\\
&\quad\times
u_{\bm{k}s\ell_3}^{c*}
u_{\bm{k}s\ell_1}^{c}
u_{\bm{k}+\bm{q}s\rq{}\ell_2}^{c*}
u_{\bm{k}+\bm{q}s\rq{}\ell_4}^{c}
\frac{
f(\varepsilon_{\bm{k}+\bm{q}s\rq{}})-f(\varepsilon_{\bm{k}s})}{
\varepsilon_{\bm{k}s}-\varepsilon_{\bm{k}+\bm{q}s\rq{}}},
\label{eq:Kq}
\end{align}
which consists of a square of the energy denominator $(\mu-\varepsilon_{\Gamma_8}^{f})^{-2}$, 
4-producted $f$-$c$ mixings 
and the $c$ band eigenvectors, 
and the Lindhard function 
with 
$\varepsilon_{\bm{k}s}$. 
Thus it has $4^4=256$ components of $f$-basis $\{m_1,m_2,m_3,m_4\}$ for each $\bm{q}$, 
and has to be summed for the $c$ orbitals $\{\ell_1,\ell_2,\ell_3,\ell_4\}$ (60$^{4}$) and the band-indexes $\{s,s\rq{}\}$ ($60^{2}$). 
Then we introduce a $f$-$c$ mixing matrix $v_{\bm{k}s}^{mm\rq{}}$ between $\{m,m\rq{}\}$ 
via the $c$ band state with $\bm{k},s$ as follows, 
\begin{align}
v_{\bm{k}s}^{mm\rq{}}=\sum_{\ell\ell\rq{}}
V_{\bm{k}m\ell}^{*}
V_{\bm{k}m\rq{}\ell\rq{}}
u_{\bm{k}s\ell}^{c*}
u_{\bm{k}s\ell\rq{}}^{c},
\end{align}
which includes whole information about the $f$ state scattering between $\{m,m\rq{}\}$ through the $c$ state with $\bm{k},s$, 
and has only $4^2=16$ components of $\{m,m\rq{}\}$ for each $\bm{k},s$ with 
a summation for $\{\ell,\ell\rq{}\}$ (60$^{2}$). 
Hence once we calculate $v_{\bm{k}s}^{mm\rq{}}$, 
$K_{m_1m_2m_3m_4}(\bm{q})$ can be easily obtained by the following compact form, 
\begin{align}
K_{m_1m_2m_3m_4}(\bm{q})=\frac{1}{N}\sum_{\bm{k}ss\rq{}}
\frac{
v_{\bm{k}s}^{m_3m_1}
v_{\bm{k}+\bm{q}s\rq{}}^{m_2m_4}
}{(\mu-\varepsilon_{\Gamma_8}^{f})^2}
\frac{
f(\varepsilon_{\bm{k}+\bm{q}s\rq{}})-f(\varepsilon_{\bm{k}s})}{
\varepsilon_{\bm{k}s}-\varepsilon_{\bm{k}+\bm{q}s\rq{}}}.
\label{eq:Kq2}
\end{align}
This expression helps us calculate all the contributions of the 60 $c$ electron charge and/or orbital fluctuations to the RKKY multipole couplings.

\begin{figure*}[t]
\centering
\includegraphics[width=14.0cm]{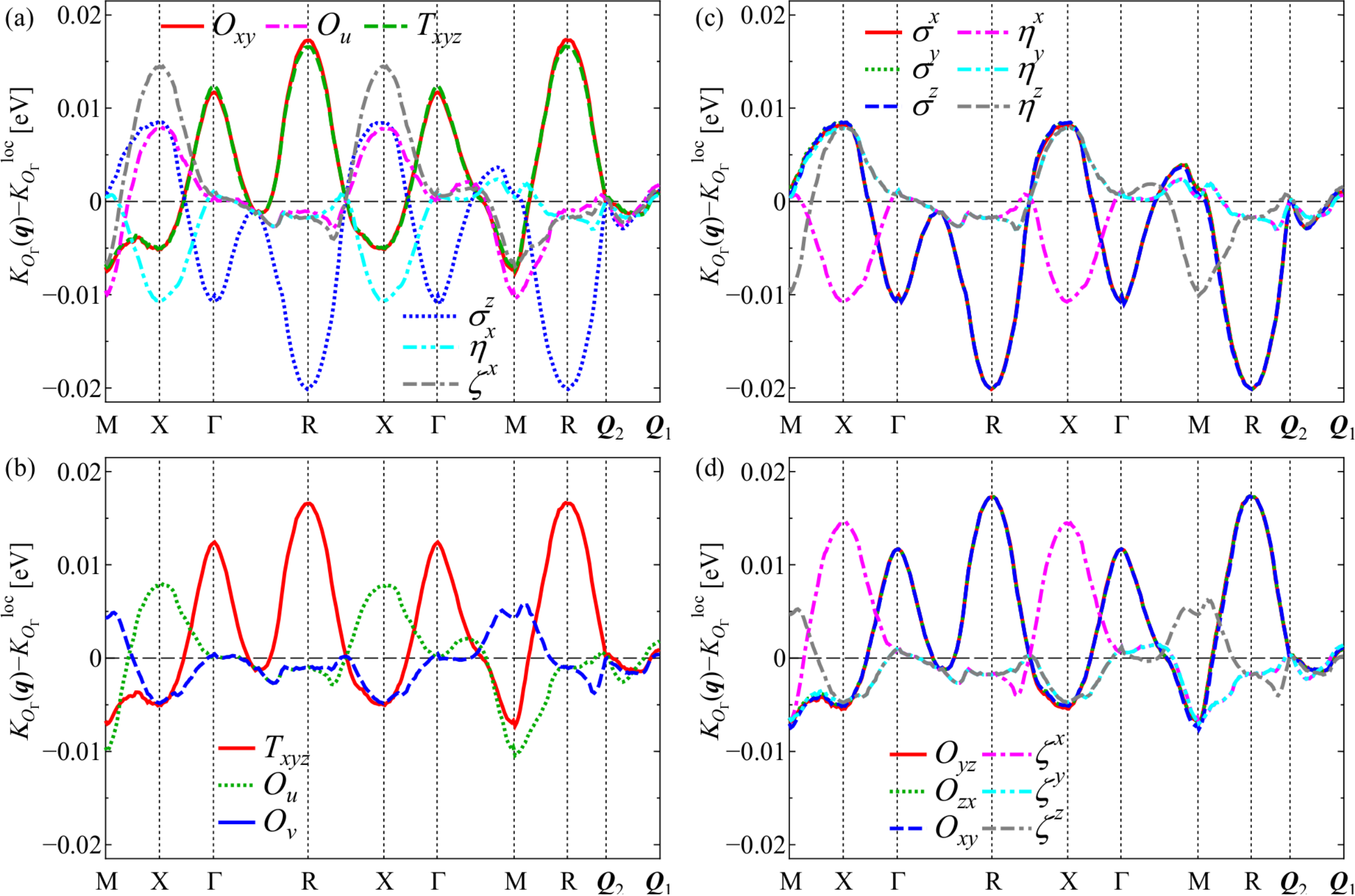}
\caption{(Color online) 
(a)-(b): The $\bm{q}$-dependence of the RKKY multipole coupling $\overline{K}_{O_{\Gamma}}^{}(\bm{q})$ 
for (a) the multipole moments of the typical irreducible representations. 
The individual terms for (b) $\Gamma_{2u}$ octupole $T_{xyz}$, $\Gamma_{3g}$ quadrupoles $O_{u},O_{v}$, 
(c) $\Gamma_{4u}$ multipoles $(\sigma^{x},\sigma^{y},\sigma^{z})$ and $(\eta^{x},\eta^{y},\eta^{z})$ and, 
(d) $\Gamma_{5g}$ quadrupoles $(O_{yz},O_{zx},O_{xy})$ and $\Gamma_{5u}$ octupoles $(\zeta^{x},\zeta^{y},\zeta^{z})$. 
$\bm{Q}_{1}=(\tfrac{\pi}{2},\tfrac{\pi}{2},0)$ and $\bm{Q}_{2}=(\tfrac{\pi}{2},\tfrac{\pi}{2},\pi)$ are the AFM ordering vectors of phase III. 
}
\label{Fig06}
\end{figure*}

In order to search the actual multipole ordering, we employ the mean-field (MF) approximation w. r. t. the multipole operator $O_{\Gamma}$, resulting in the MF Hamiltonian as follows, 
\begin{align}
&H_{\rm RKKY}^{\rm MF}=-\sum_{\bm{q}}\sum_{\Gamma}\overline{K}_{O_{\Gamma}}(\bm{q})
\overline{O}_{\Gamma}(\bm{q})\overline{O}_{\Gamma}(-\bm{q}),\label{eq:HMO}
\end{align}
where the multipole coupling $\overline{K}_{O_{\Gamma}}(\bm{q})=K_{O_{\Gamma}}(\bm{q})-K_{O_{\Gamma}}^{\rm loc}$ 
and 
$K_{O_{\Gamma}}(\bm{q})$ and the MF order parameter $\overline{O}_{\Gamma}(\bm{q})$ 
are given by,
\begin{align}
&K_{O_{\Gamma}}(\bm{q})
=\sum_{m_1m_2}\sum_{m_3m_4}O_{m_1m_2}^{\Gamma}O_{m_4m_3}^{\Gamma}K_{m_1m_2m_3m_4}(\bm{q}),\\
&\overline{O}_{\Gamma}(\bm{q})=\frac{1}{N}\sum_{i}\overline{O}_{\Gamma}(\bm{R}_i)e^{i\bm{q}\cdot\bm{R}_{i}},
\end{align}
where $K_{O_{\Gamma}}^{\rm loc}=(1/N)\sum_{\bm{q}}K_{O_{\Gamma}}(\bm{q})$ and 
$\overline{O}_{\Gamma}(\bm{R}_i)$ is MF multipole order parameter defined at the unit-cell vector $\bm{R}_{i}$. 
Then the MF multipole susceptibility $\chi_{O_{\Gamma}}^{\rm MF}(\bm{q})$ is written by,
\begin{align}
\chi_{O_{\Gamma}}^{\rm MF}(\bm{q})=\frac{
\chi_{O_{\Gamma}}^{}(\bm{q})}{1-\chi_{O_{\Gamma}}^{}(\bm{q})\overline{K}_{O_{\Gamma}}(\bm{q})},
\label{eq:chi_OMF}
\end{align}
which is enhanced towards the multipole ordering instability for the ordering moment $O_{\Gamma}$ and wavevector $\bm{q}$, 
and finally diverges at a critical point of the multipole ordering transition temperature $T=T_{O_{\Gamma}}$ 
where $\chi_{O_{\Gamma}}^{}(\bm{q})\overline{K}_{O_{\Gamma}}(\bm{q})$ reaches unity. 
The $\bm{q}$-dependence of $\chi_{O_{\Gamma}}^{}(\bm{q})$ is weak as shown in Sec. \ref{sec3}, 
and then the sign and maximum value of $\overline{K}_{O_{\Gamma}}^{}(\bm{q})$ 
determines the multipole ordering moment and wavevector for any given $T$. 
Hereafter we set $\mu-\varepsilon_{\Gamma_8}^{f}=1$ eV for simplicity, 
since this factor is independent of $\bm{q}$ and $m$, and hence does no affect the ordering type and wavevector, 
whose effect is discussed in Sec. \ref{sec5}.

\subsection{$\bm{q}$-dependence of RKKY coupling $\overline{K}_{O_{\Gamma}}^{}(\bm{q})$}
The obtained RKKY multipole couplings $\overline{K}_{O_{\Gamma}}^{}(\bm{q})$ for several multipole moments along the high symmetry line in the BZ are plotted as shown in Figs. \ref{Fig06} (a)-(d), 
where the positive (negative) coupling for a certain multipole $O_{\Gamma}$ and wavevector $\bm{q}$ 
enhances (suppresses) the corresponding multipole fluctuation 
as explained in Eq. (\ref{eq:chi_OMF}), and its positive maximum value gives a leading multipole ordering mode. 
The obtained results for the leading multipole ordering modes upto the 10th largest coupling
are summarized in Table \ref{table02}. 

\begin{table}[t]
\centering
\scalebox{0.84}{
\begin{tabular}{cccccc}
\hline \rule{0pt}{4mm}
rank & IRR & multipole & wavevector & value [meV] & ratio  \tabularnewline \hline\hline \rule{0pt}{4mm}
1 & $\Gamma_{5g}$ & $(O_{yz},O_{zx},O_{xy})$ & $(\pi,\pi,\pi)$ & 17.26 & 1.00 \rule{0pt}{5mm}\tabularnewline[2pt] \hdashline \rule{0pt}{5mm}
2 & $\Gamma_{2u}$ & $T_{xyz}$ & $(\pi,\pi,\pi)$ & 16.56 & 0.96 \rule{0pt}{5mm}\tabularnewline[2pt] \hline \rule{0pt}{8mm}
3 & $\Gamma_{5u}$ & \pb{$\zeta^{x}$\\ $\zeta^{y}$\\ $\zeta^{z}$} & \pb{$(\pi,0,0)$\\ $(0,\pi,0)$\\ $(0,0,\pi)$} & 14.48 & 0.84 \rule{0pt}{8mm}\tabularnewline[15pt] \hdashline \rule{0pt}{8mm}
4 & $\Gamma_{3g}$ & \pb{$O_{y^2-z^2}$\\ $O_{z^2-x^2}$\\ $O_{x^2-y^2}$} & \pb{$(\pi,0,0)$\\ $(0,\pi,0)$\\ $(0,0,\pi)$} & 14.08 & 0.82 \rule{0pt}{8mm}\tabularnewline[14pt] \hline \rule{0pt}{4mm}
5 & $\Gamma_{2u}$ & $T_{xyz}$ & $(0,0,0)$ & 12.43 & 0.72 \rule{0pt}{5mm}\tabularnewline[2pt] \hdashline \rule{0pt}{5mm}
6 & $\Gamma_{5g}$ & $(O_{yz},O_{zx},O_{xy})$ & $(0,0,0)$ & 11.69 & 0.68 \rule{0pt}{5mm}\tabularnewline[2pt] \hline \rule{0pt}{8mm}
7  & $\Gamma_{4u}^{(1)}$ & \pb{$\sigma^{x}$\\ $\sigma^{y}$\\ $\sigma^{z}$} & \pb{$(0,\pi,0),(0,0,\pi)$\\ $(0,0,\pi),(\pi,0,0)$\\ $(\pi,0,0),(0,\pi,0)$} & 8.41 & 0.49 \rule{0pt}{1mm}\tabularnewline[14pt] \hline \rule{0pt}{8mm}
8  & $\Gamma_{4u}^{(1)}$ & \pb{$\sigma^{x}$\\ $\sigma^{y}$\\ $\sigma^{z}$} & \pb{$(\pi,0,0)$\\ $(0,\pi,0)$\\ $(0,0,\pi)$} & 8.11 & 0.47 \rule{0pt}{1mm}\tabularnewline[14pt] \hline \rule{0pt}{8mm}
9  & $\Gamma_{4u}^{(2)}$ & \pb{$\eta^{x}$\\ $\eta^{y}$\\ $\eta^{z}$} & \pb{$(0,\pi,0),(0,0,\pi)$\\ $(0,0,\pi),(\pi,0,0)$\\ $(\pi,0,0),(0,\pi,0)$} & 7.87 & 0.46 \rule{0pt}{1mm}\tabularnewline[14pt] \hdashline \rule{0pt}{8mm}
10 & $\Gamma_{3g}$ & \pb{$O_{3x^2-r^2}$\\ $O_{3y^2-r^2}$\\ $O_{3z^2-r2}$} & \pb{$(0,\pi,0),(0,0,\pi)$\\ $(0,0,\pi),(\pi,0,0)$\\ $(\pi,0,0),(0,\pi,0)$} & 7.78 & 0.45 \rule{0pt}{1mm}\tabularnewline[14pt]
\hline
\end{tabular}
}
\caption{
The obtained possible multipole modes upto the 10th largest coupling together with 
the corresponding moment types, wavevectors, maximum values and ratios to the largest value of $\Gamma_{5g}$-$(\pi,\pi,\pi)$, 
where the $\Gamma_{3g}$ moments $O_{y^2-z^2}$, $O_{z^2-x^2}$, $O_{3x^2-r^2}$ and $O_{3y^2-r^2}$ 
are described by the linear combinations of $O_{u}(=O_{3z^2-r^2})$ and $O_{v}(=O_{x^2-y^2})$. 
}\label{table02}
\end{table} 

The couplings of the $\Gamma_{5g}$ quadrupoles $(O_{yz},O_{zx},O_{xy})$ for $\bm{q}=(\pi,\pi,\pi)$ become largest among all moments and $\bm{q}$, which perfectly corresponds to the AFQ ordering of CeB$_6$. 
In addition, $\Gamma_{2u}$ ocutupole $T_{xyz}$ coupling is quite large and comparable to the $\Gamma_{5g}$ quadrupoles with the same wavevector as shown in Fig. \ref{Fig06} (a) but slightly small within the present calculation accuracy as shown in Table \ref{table02}, 
which seems to be the same value from the previous discussions\cite{Shiba1999,Hanzawa2000} 
where $O_{xy}$ and $T_{xyz}$ have almost same matrix elements and yield the similar fluctuations in phase I. 
Furthermore the quadupoles $(O_{yz},O_{zx},O_{xy})$ and octupole $T_{xyz}$ couplings 
also take a substantial peak for $\bm{q}=(0,0,0)$ as shown in Figs. \ref{Fig06} (a) and (d) 
and correspond to the elastic softening of $C_{44}$\cite{Luthi1984,Goto1985,Nakamura1994,Nakamura1995}. 

The next largest coupling is the $\Gamma_{5u}$ octupole $\zeta^{x}$ at $\bm{q}=(\pi,0,0)$ [Figs. \ref{Fig06} (a) \& (d)]
which is degenerate for $\zeta^{y}~[\zeta^{z}]$ octupole at $\bm{q}=(0,\pi,0)~[(0,0,\pi)]$ due to the cubic symmetry. 
The role of the $\Gamma_{5u}$ octupoles $(\zeta^{x},\zeta^{y},\zeta^{z})$ is also discussed for the phase IV observed in the La-doping system Ce$_x$La$_{1-x}$B$_6$ with $x<0.8$
\cite{Kubo-Kuramoto2003,Kubo-Kuramoto2004,Mannix2005,Kurasawa2007,Matsumura2014,Inami2014,Sera2018}, 
where the $\bm{q}=(\pi,\pi,\pi)$ antiferro-octupolar (AFO) ordering of $(\zeta^{x}+\zeta^{y}+\zeta^{z})/\sqrt{3}$ is considered to be a possible mode. In contrast, the present theory suggests 
the $\Gamma_{5u}$ AFO with the domained structure of $\zeta^{x}$, $\zeta^{y}$ and $\zeta^{z}$
for $\bm{q}=(\pi,0,0)$, $(0,\pi,0)$ and $(0,0,\pi)$, respectively, and this point will be discussed in the next subsection. 

In addition to this, the $\Gamma_{3g}$ quadrupole $O_{v}=O_{x^2-y^2}$ coupling is quite large for $\bm{q}=(0,0,\pi)$ (not shown) 
and becomes similar value of the octupole coupling $\zeta^{z}$ as shown in Table \ref{table02}, 
which is also degenerate for the rotated moments to the each principle-axis $O_{y^2-z^2}$ and $O_{z^2-x^2}$. 
This is namely the $\Gamma_{3g}$-AFQ mode where 
the moment directions and wavevectors 
are perpendicular such as the multipole moments of $O_{y^2-z^2}$, $O_{z^2-x^2}$ and $O_{x^2-y^2}$ with the corresponding wavevectors for $\bm{q}=(\pi,0,0)$, $(0,\pi,0)$ and $(0,0,\pi)$ respectively. 

The $\Gamma_{4u}$ magnetic multipole couplings of $(\sigma^{x},\sigma^{y},\sigma^{z})$ and $(\eta^{x},\eta^{y},\eta^{z})$ 
are plotted in Fig. \ref{Fig04} (d) and their maximum values in $\bm{q}$-space 
are smaller than that of the quadrupole and octupole couplings as shown in Table \ref{table02}, 
where the $\Gamma_{4u}$ maximum peak values are less than half of the first leading peak value of the $\Gamma_{5g}$-$(\pi,\pi,\pi)$. 
At the AFM ordering vectors for phase III, $\bm{Q}_1$ and $\bm{Q}_{2}$, 
the couplings of the magnetic multipoles $(\sigma^{x},\sigma^{y},\sigma^{z})$ have small peaks as shown in Fig. \ref{Fig04} (e) and they shall be enhanced and dominant only when the system enters into phase II, which is not discussed in the present paper. 

As usually discussed in the itinerant $f$ electron picture with the multi-orbital Hubbard model\cite{Ikeda2012,Nomoto-Ikeda2014,Nomoto-Ikeda2016}, 
the weak coupling theory like the RPA and its extensions yields largely enhanced magnetic multipole (spin) fluctuations 
which become always larger than the nonmagnetic multipole (orbital) fluctuations like $\Gamma_{5g}$ multipoles here. 
As for CeB$_6$, the $f$ electron itself is already localized at each Ce site 
and the remained magnetic and nonmagnetic multipole moments interact with the RKKY intersite couplings, 
where the magnetic multipole coupling does not necessarily dominate over the nonmagnetic one, 
since the dominant RKKY coupling is determined by the detail of the $f$-$c$ mixing and the meditating $c$ electron charge and/or orbital fluctuations. 

Here we note the $c$ electron charge and orbital fluctuations and their contribution to the coupling $\overline{K}_{O_{\Gamma}}^{}(\bm{q})$. 
By changing summation for the $c$ orbital-set in Eq. (\ref{eq:Kq}) and the band-index in Eq. (\ref{eq:Kq2}), 
we have obtained that 
both effects of 
the Ce-$d_{eg}$ orbitals of $d_{x^2-y^2}$ and $d_{3z^2-r^2}$ distributed in the 21st and 23rd bands 
and 
the charge fluctuation of B$_6$-molecule having a maximum at $\bm{q}=(\pi,\pi,\pi)$ and large values along R-M line 
play significant roles for the $\Gamma_{5g}$ AFQ mode. 
In particular, we observe a non-negligible contribution from the 23rd band 
which does not have FS but is very close to $E_{\rm F}$ along the $\Gamma$-M direction as shown in Fig. \ref{Fig05} (b). 
The explicit results and further analysis of such $c$ electron contributions to the multipole couplings will be presented in elsewhere.


\begin{figure*}[t]
\begin{minipage}{0.7\hsize}
\begin{center}
\includegraphics[width=11.0cm]{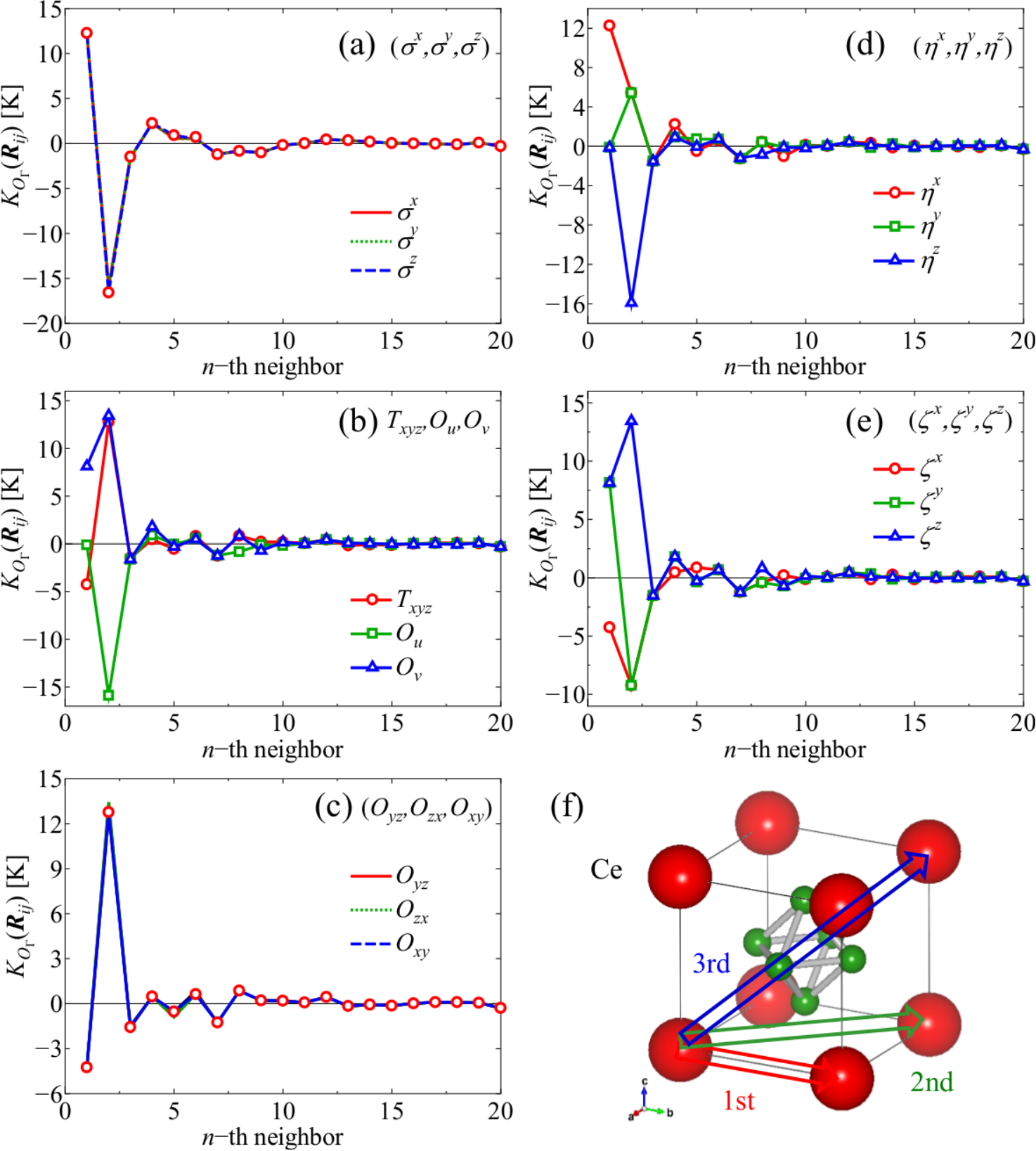}
\caption{(Color online) 
The $n$-neighbor intersite vector $\bm{R}_{ij}$-dependence of the RKKY multipole coupling $\overline{K}_{O_{\Gamma}}^{}(\bm{R}_{ij})$ for 
(a) $\Gamma_{4u}^{(1)}$ multipoles $(\sigma^{x},\sigma^{y},\sigma^{z})$ (magnetic moments), 
(b) $\Gamma_{2u}$ octupole $T_{xyz}$ and $\Gamma_{3g}$ quadrupoles ($O_{u},O_{v})$, 
(c) $\Gamma_{5g}$ quadrupoles $(O_{yz},O_{zx},O_{xy})$, 
(d) $\Gamma_{4u}^{(2)}$ multipoles $(\eta^{x},\eta^{y},\eta^{z})$, and 
(e) $\Gamma_{5u}$ octupoles $(\zeta^{x},\zeta^{y},\zeta^{z})$, 
where the ferro (antiferro) coupling corresponds to the positive (negative) sign. 
(f) The first (red), second (green) and third (blue) neighbor intersite vectors between Ce-atoms in CeB$_6$. 
}
\label{Fig07}
\end{center}
\end{minipage}
  \def\@captype{table}
\begin{minipage}{0.3\hsize}
\begin{center}
\scalebox{0.89}{
\begin{tabular}{cccc}
\hline
$n$-th & $(n_1,n_2,n_3)$ & $|\bm{R}_{ij}|/a$ \\ \hline\hline
 1 & (1,0,0) & $1$ \\
 2 & (1,1,0) & $\sqrt{2}$ \\ 
 3 & (1,1,1) & $\sqrt{3}$ \\
 4 & (2,0,0) & $\sqrt{4}$ \\
 5 & (2,1,0) & $\sqrt{5}$ \\
 6 & (2,1,1) & $\sqrt{6}$ \\
 7 & (2,2,0) & $\sqrt{8}$ \\
 8 & (2,2,1) & $\sqrt{9}$ \\
 9 & (3,0,0) & $\sqrt{9}$ \\
10 & (3,1,0) & $\sqrt{10}$ \\
11 & (3,1,1) & $\sqrt{11}$ \\
12 & (2,2,2) & $\sqrt{12}$ \\
13 & (3,2,0) & $\sqrt{13}$ \\
14 & (3,2,1) & $\sqrt{14}$ \\
15 & (4,0,0) & $\sqrt{16}$ \\
16 & (4,1,0) & $\sqrt{17}$ \\
17 & (3,2,2) & $\sqrt{17}$ \\
18 & (4,1,1) & $\sqrt{18}$ \\
19 & (3,3,0) & $\sqrt{18}$ \\
20 & (3,3,1) & $\sqrt{19}$ \\
\hline
\end{tabular}
}
\tblcaption{
The $n$-th neighbor sites and distances for the intersite vectors between Ce-Ce 
$\bm{R}_{ij}=a\left(n_1\bm{e}_x+n_2\bm{e}_y+n_3\bm{e}_z
\right)$ where $\{\bm{e}_x,\bm{e}_y,\bm{e}_z\}$ are unit vectors along $x,y,z$-direction. 
}\label{table03}
\end{center}
\end{minipage}
\end{figure*}

\subsection{$\bm{R}_{ij}$-dependence of RKKY coupling $K_{O_{\Gamma}}^{}(\bm{R}_{ij})$}
In general, the RKKY interaction is known to have long-range and oscillating features discussed in the early studies\cite{SL1985,Frisken-Miller1986}. 
For the multipole ordering of CeB$_6$, however, the coupling is limited only in the nearest neighbor terms in the previous studies\cite{Ohkawa1983,Ohkawa1985,Shiina1997,Shiina1998,Shiba1999,Hanzawa2000,Sakurai-Kuramoto2004,Sakurai2005}. 
In contrast, the present formalism provides the real space dependent couplings $K_{O_{\Gamma}}^{}(\bm{R}_{ij})$ 
which is explicitly written as
\begin{align}
&K_{O_{\Gamma}}^{}(\bm{R}_{ij})
=\sum_{m_1m_2}\sum_{m_3m_4}O_{m_1m_2}^{\Gamma}O_{m_4m_3}^{\Gamma}K_{m_1m_2m_3m_4}(\bm{R}_{ij}),
\end{align}
where $K_{m_1m_2m_3m_4}(\bm{R}_{ij})$ is given in Eq. (\ref{eq:Kij}).

Figure \ref{Fig07} shows the site-dependence of the RKKY multipole couplings $K_{O_{\Gamma}}^{}(\bm{R}_{ij})$ 
with the intersite vector $\bm{R}_{ij}$ upto 20-th neighbor sites as shown in Table \ref{table03}, 
where the positive (negative) sign corresponds to the ferro (antiffero) coupling for each neighboring site. 

As shown in Fig. \ref{Fig07} (a), the $\Gamma_{4u}^{(1)}$ magnetic multipole $(\sigma^{x},\sigma^{y},\sigma^{z})$ couplings 
exhibit several sign changes with a few site-intervals and 
degeneracy due to the symmetry of paramagnetic phase for all $n$-th neighbors, 
where $\sigma^{x}$, $\sigma^{y}$ and $\sigma^{z}$ corresponds to the $x,~y$- and $z$-moment direction respectively. 
We also confirm that a monopole operator $I$ defined as a unit matrix for $\Gamma_{8}$-basis 
is also degenerate with $(\sigma^{x},\sigma^{y},\sigma^{z})$ possessing the same oscillating feature.  
The couplings of $I$ and $(\sigma^{x},\sigma^{y},\sigma^{z})$ are isotropic, 
where for example they have the same value for 6 first neighbor sites $\bm{R}_{ij}=(\pm a,0,0)$, $(0,\pm a,0)$ and $(0,\pm a,0)$. 

The $\Gamma_{5g}$ and $\Gamma_{2u}$ multipoles couplings which gives the leading AFQ mode 
show staggered and isotropic behaviors, 
where the first, second and third neighbor couplings show positive, negative and positive signs respectively
as shown in Figs. \ref{Fig07} (b) [$T_{xyz}$] and (c) [$(O_{yz},O_{zx},O_{xy})$], which clearly enhance cooperatively the antiferro ordering. 
In particular, the second neighbor coupling becomes largest with positive sign, 
which also enhances the AFQ mode as a main driving force. 

On the other hands, the main origin of the $(\pi,0,0)$-AFO with $\Gamma_{5u}$ octupole $\zeta^{x}$ 
is the anisotropic first neighbor couplings as shown in Fig. \ref{Fig05} (e), 
where the coupling of which the intersite vector and moment direction are parallel (perpendicular) each other 
has negative (positive) sign such as $K^{x}_{{\rm 5}u}(\bm{R}_{x})<0$ and $K^{y}_{{\rm 5}u}(\bm{R}_{x}),~K^{z}_{{\rm 5}u}(\bm{R}_{x})>0$ for $\bm{R}_{x}=(a,0,0)$, 
which yields $K^{x}_{{\rm 5}u}(\bm{R}_{y}),~K^{x}_{{\rm 5}u}(\bm{R}_{z})>0$ for the perpendicular first neighbors $\bm{R}_{y}=(0,a,0)$ and $\bm{R}_{z}=(0,0,a)$, resulting in the enhancement of the $\bm{q}=(\pi,0,0)$ mode by $K^{x}_{{\rm 5}u}(\bm{q})=2K^{x}_{{\rm 5}u}(\bm{R}_{x})\cos q_{x}+2K^{x}_{{\rm 5}u}(\bm{R}_{y})\cos q_{y}+2K^{x}_{{\rm 5}u}(\bm{R}_{z})\cos q_{z}$. 

With this matter in mind, 
we might be able to explain the doping phase diagram of Ce$_x$La$_{1-x}$B$_6$, 
where the doping effect is simply treated by the reduction of the multipole coupling depending the coordination number for each site, since a La-substituted site has no multipole moment. 
Consequently, the first leading $(\pi,\pi,\pi)$-AFQ mode in $x=1$ decreases with decreasing $x$ more rapidly than 
the second $(\pi,0,0)$-AFO mode due to the difference of the coordination numbers : 
6 first neighbors and 12 second neighbors for $(\pi,\pi,\pi)$, 
while 2 parallel first neighbors and 4 perpendicular first neighbors and so on for $(\pi,0,0)$. 
This turnover of the dominant mode may be consistent with the recent inelastic neutron scattering (INS) experiments in the La-doping system\cite{Nikitin2018} 
where the $\bm{q}=(\pi,0,0)$ intensity is developed and becomes dominant mode for $x<0.8$. 

As well as the $\Gamma_{5u}$ couplings, the $\Gamma_{3g}$ quadrupoles $(O_{u},O_{v})$ and $\Gamma_{4u}^{(2)}$ multipole $(\eta^{x},\eta^{y},\eta^{z})$ couplings exhibit the anisotropic behavior as shown in Figs. \ref{Fig07} (b) and (d). 
The obtained results of such long-range, oscillating, isotropic or anisotropic behaviors depending 
on the multipole moments seem to be worthwhile to study from now on. 


\section{Summary and Discussion}\label{sec5}
In summary, we study the electronic states of CeB$_6$ and perform a direct calculation of the RKKY interaction based on the 74-orbital effective Wannier model derived from the bandstructure calculation and obtain the following results. 

(1) When the $f$-$f$ hopping and $f$-$c$ mixing of the Wannier model are suppressed by the renormalization factor $Z_{f}$ based on the FL theory, 
the quasi-particle band states are observed where 
the fully dispersionless $f$ bands slightly hybridize with the wide-bandwidth $c$ bands, 
which is almost the same as the GGA+$U$-band of LaB$_6$ 
having a single ellipsoidal FS centered at X point. 
This is in good agreement with the recent ARPES\cite{Souma2004,Neupane2015,Ramankuttya2016,Koitzsch2016} and early dHvA experiments\cite{Onuki1989,Endo2006} and strongly supports the localized $f$ electron picture for CeB$_6$. 

(2) By using the LaB$_6$-like $c$ band states together with the $f$-$c$ mixing elements of the CeB$_6$ Wannier model,  
we calculate the RKKY couplings for all active multipole moments in $\Gamma_8$ subspace explicitly
as functions of wavevector $\bm{q}$ and inter-cell vector $\bm{R}_{ij}$, 
where we derive a useful expression in order to treat all 60 $c$-orbital contributions. 

(3) The couplings of the $\Gamma_{5g}$ quadrupole $(O_{yz},O_{zx},O_{xy})$ together with the $\Gamma_{2u}$ octupole $T_{xyz}$ are highly enhanced for $\bm{q}=(\pi,\pi,\pi)$ as the 1st and 2nd leading modes, and $\bm{q}=(0,0,0)$ as the 5th and 6th leading modes, 
where the former explains the AFQ ordering of the phase II and the latter corresponds to the elastic softening of $C_{44}$. 
The 3rd (4th) leading mode is the $\Gamma_{5u}$-AFO ($\Gamma_{3g}$-AFQ) of $\zeta^{x}$, $\zeta^{y}$ and $\zeta^{z}$ ($O_{y^2-z^2}$, $O_{z^2-x^2}$ and $O_{x^2-y^2}$) with the corresponding wavevectors for $\bm{q}=(\pi,0,0)$, $(0,\pi,0)$ and $(0,0,\pi)$ respectively, 
which are almost degenerate each other and differ from the discussed AFO-mode with $(\zeta^{x}+\zeta^{y}+\zeta^{z})/\sqrt{3}$ at $\bm{q}=(\pi,\pi,\pi)$\cite{Kubo-Kuramoto2003,Kubo-Kuramoto2004,Mannix2005,Kurasawa2007,Matsumura2014,Inami2014,Sera2018}. 

(4) All the obtained RKKY couplings have long-range and oscillating behavior as a function of $\bm{R}_{ij}$, 
where the $\Gamma_{5g}$ quadrupole $(O_{yz},O_{zx},O_{xy})$ and $\Gamma_{2u}$ octupole $T_{xyz}$ couplings 
indicate the sign-reversing for each neighboring site and 
have a positive largest value at the second neighbor which cooperatively enhances the AFQ with $\bm{q}=(\pi,\pi,\pi)$, 
while for the second leading $\Gamma_{5u}$ AFO mode, the anisotropic first neighbor couplings are significant. 
This induces the leading mode shift with increasing the La-substitution rate $x$ in Ce$_x$La$_{1-x}$B$_6$
from the $(\pi,\pi,\pi)$-AFQ with $O_{xy}$ (phase II) to the $(\pi,0,0)$-AFO with $\zeta^{x}$ (phase IV)  
which may be also consistent with the $(\pi,0,0)$ peak in the INS data\cite{Nikitin2018}. 

(5) The present approach can determine the possible type of the multipole moment and the ordering vector $\bm{q}$ definitely 
once the $c$ band states and $f$-$c$ mixings are given by the bandstructure calculation, 
which enables us to discuss the inherent feature and the concrete situation of actual compounds 
such as the changes of FSs, carrier densities, lattice constants and internal coordinates of atoms. 

In this study, we take only the $f^{0}$-process and assume that the contribution from $f^{2}$-process is the same as that of $f^{0}$, since the $f^{0}$-process is fully one-body effect and directly obtained from the DFT-bandstructure calculation. 
As mentioned in Sec. \ref{sec4}, we take $\mu-\varepsilon_{\Gamma_8}^{f}=1$ eV, 
but the excitation energy from the $f^1$-stable to $f^0$-intermediate states in CeB$_6$ 
is roughly estimated by $\mu-\varepsilon_{\Gamma_8}^{f}=2\sim4$ eV\cite{Neupane2015,Koitzsch2016}, 
so that our results of $\overline{K}_{O_{\Gamma}}$ obtained in Sec. \ref{sec4} should be multiplied by 
a single reduction factor $\frac{1}{4}\sim\frac{1}{16}$, which 
yields the same order of the actual transition temperature of CeB$_6$ as a few K, 
for example, the inter-quadrupole coupling value $K_{\Gamma_{5g}}\rq{}=2.1$ K\cite{Nakamura1994}. 

The explicit determination of the couplings and the transition temperatures needs the many-body energy difference between the ground and intermediate $f^{0}$ and $f^{2}$ states, 
where to what extent the many-body effect from the $f^{2}$ and more multiple $f$ processes changes the present result is an important question elucidating the multipole ordering system with different valence materials such as PrB$_6$ and NdB$_6$. 
The explicit calculation of the coupling including the $f^{2}$-process and/or more many-body contribution, 
and the whole phase diagram in $T$-$H$ plane will be presented in the subsequent paper. 

As a complementary approach to the present localized $f$ electron treatment, 
the dynamical mean field theory\cite{Georges1996} 
enabling to take account of the full local correlation effect
and its extensions\cite{Rohringer2018} including the intersite correlation 
could be valid for directly describing the fully localized $f$ states starting from the itinerant $f$ states 
and their multiple ordering phenomena including superconductivity\cite{Otsuki2015}. 
The application of such many-body theory to the realistic materials and their comparison with present theory are also the essential future problems.

\begin{acknowledgment}
We would like to thank Y. {\=O}no and Y. Iizuka for valuable comments and discussions.
\end{acknowledgment}


\bibliography{CeB6-RKKY_cmat}

\providecommand{\noopsort}[1]{}\providecommand{\singleletter}[1]{#1}%
\begin{thebibliography}{10}

\bibitem{Santini2009}
P.~Santini, S.~Carretta, G.~Amorett, R.~Caeiuffo, N.~Mabnani, and G.~H. Lander:
  Rev. Mod. Phys. {\bfseries 81} (2009) 807.

\bibitem{Kuramoto2009}
Y.~Kuramoto, H.~Kusunose, and A.~Kiss: J. Phys. Soc. Jpn. {\bfseries 78} (2009)
  072001.

\bibitem{Kusunose2008}
H.~Kusunose: J. Phys. Soc. Jpn. {\bfseries 77} (2008) 064710.

\bibitem{Cameron2016}
A.~S. Cameron, G.~Friemel, and D.~S. Inosov: Rev. Prog. Phys. {\bfseries 79}
  (2016) 066502.

\bibitem{Fujita1980}
T.~Fujita, M.~Suzuki, T.~Komatsubara, S.~Kunii, T.~Kasuya, and T.~Ohtsuka:
  Solid State Commun. {\bfseries 35} (1980) 569.

\bibitem{Kawakami1980}
M.~Kawakami, S.~Kunii, T.~Komatsubara, and T.~Kasuya: Solid State Commun.
  {\bfseries 36} (1980) 435.

\bibitem{Takase1980}
A.~Takase, K.~Kojima, and T.~K. andT. Kasuya: Solid State Commun. {\bfseries
  36} (1980) 461.

\bibitem{Zirngiebl1984}
E.~Zirngiebl, B.~Hillebrands, S.~Blumenr$\ddot{\rm o}$der, G.~G$\ddot{\rm
  u}$ntherodt, M.~Loewenhaupt, J.~M. Carpenter, K.~Winzer, and Z.~Fisk: Phys.
  Rev. B {\bfseries 30} (1984) 4052.

\bibitem{Furuno1985}
T.~Furuno, N.~Sato, S.~Kunii, T.~Kasuya, and W.~Sakai: J. Phys. Soc. Jpn.
  {\bfseries 54} (1985) 1899.

\bibitem{Bredl1987}
C.~D. Bredl: J. Magn. Magn. Mater. {\bfseries 63-64} (1987) 355.

\bibitem{Effantin1982}
J.~M. Effantin, P.Burlet, J.Rossat-Mignod, S.Kunii, and T.Kasuya: {\em Valence
  Instabilities} (North-Holland Publishing Company, ed. P. Wachter and H.
  Boppart, 1982), p. 559.

\bibitem{Effantin1985}
J.~M. Effantin, J.~Rossat-Mignod, P.~Burlet, H.~Bartholin, S.~Kunii, and
  T.~Kasuya: J. Magn. Magn. Mater. {\bfseries 47-48} (1985) 145.

\bibitem{Erkelens1987}
W.~A.~C. Erkelens, L.~P. Regnault, P.~Burlet, J.~Rossat-Mignod, S.~Kunii, and
  T.~Kasuya: J. Magn. Magn. Mater. {\bfseries 63-64} (1987) 61.

\bibitem{Takigawa1983}
M.~Takigawa, H.~Yasuoka, T.~Tanaka, and Y.~Ishizawa: J. Phys. Soc. Jpn.
  {\bfseries 53} (1983) 728.

\bibitem{Luthi1984}
B.~L$\ddot{\rm u}$thi, S.~Blumenr$\ddot{\rm o}$der, B.~Hillbrands,
  E.~Zirngiebl, G.~G$\ddot{\rm u}$untherodt, and K.~Winzer: Z. Phys. B
  {\bfseries 58} (1984) 31.

\bibitem{Goto1985}
T.~Goto, A.~Tamaki, T.~Suzuki, S.~Kunii, N.~Sato, T.~Suzuki, H.~Kitazawa,
  T.~Fujimura, and T.~Kasuya: J. Magn. Magn. Mater. {\bfseries 52} (1985) 253.

\bibitem{Nakamura1994}
S.~Nakamura, T.~Goto, S.~Kunii, K.~Iwashita, and A.~Tamaki: J. Phys. Soc. Jpn.
  {\bfseries 63} (1994) 623.

\bibitem{Nakamura1995}
S.~Nakamura, T.~Goto, and S.~Kunii: J. Phys. Soc. Jpn. {\bfseries 64} (1995)
  3941.

\bibitem{Tayama1997}
T.~Tayama, T.~S.~K. Tenya, H.~Amitsuka, and S.~Kunii: J. Phys. Soc. Jpn.
  {\bfseries 66} (1997) 2268.

\bibitem{Hiroi1997}
M.~Hiroi, S.~Kobayashi, M.~Sera, N.~Kobayashi, and S.~Kunii: J. Phys. Soc. Jpn.
  {\bfseries 66} (1997) 132.

\bibitem{Sakai1997}
O.~Sakai, R.~Shiina, H.~Shiba, and P.~Thalmeier: J. Phys. Soc. Jpn. {\bfseries
  66} (1997) 3005.

\bibitem{Friemel2012}
G.~Friemel, Y.~Li, A.~Dukhnenko, N.~Shitsevalova, N.~Shuchanko, A.~Ivanov,
  V.~Filipov, B.~Keimer, and D.~Inosov: Nat. Commun. {\bfseries 3} (2012) 830.

\bibitem{Jang2014}
H.~Jang, G.~Friemel, J.~Ollivier, A.~V. Dunkhnenko, N.~Y. Shitsevalvoa, V.~B.
  Filipov, B.~Keimer, and D.~S. Inosov: Nat. Mater. {\bfseries 13} (2014) 682.

\bibitem{Jang2017}
D.~Jang, P.~Y. Portnichenko, A.~S. Cameron, G.~Friemel, A.~V. Duknenko, N.~Y.
  Shitsevalova, V.~B. Filipov, A.~Schneidewind, A.~Ivanov, D.~S. Inosov, and
  M.~Brando: npj Quantum Mater. {\bfseries 2} (2017) 62.

\bibitem{Nikitin2018}
S.~E. Nikitin, P.~Y. Portnichenko, A.~V. Dukhnenko, N.~Y. Shitsevalova, V.~B.
  Filipov, Y.~Qiu, J.~A. Rodriguez-Rivera, J.~Ollivier, and D.~S. Inosov: Phys.
  Rev. B {\bfseries 97} (2018) 075116.

\bibitem{Onuki1989}
Y.~{\=O}nuki, T.~Komatsubara, P.~H.~P. Reinders, and M.~Springford: J. Phys.
  Soc. Jpn. {\bfseries 58} (1989) 3698.

\bibitem{Endo2006}
M.~Endo, S.~Nakamura, T.~Isshiki, N.~Kimura, T.~Nojima, H.~Aoki, H.~Harima, and
  S.~Kunii: J. Phys. Soc. Jpn. {\bfseries 75} (2006) 114704.

\bibitem{Souma2004}
S.~Souma, Y.~Iida, T.~Sato, T.~Takahashi, and S.~Kunii: Physica B {\bfseries
  351} (2004) 283.

\bibitem{Neupane2015}
M.~Neupane, N.~Alidoust, I.~Belopolski, G.~Bian, S.-Y. Xu, D.-J. Kim, P.~P.
  Shibayev, D.~S. Sanchez, H.~Zheng, T.-R. Chang, H.-T. Jeng, P.~S.
  Riseborough, H.~Lin, A.~Bansil, T.~Durakiewicz, Z.~Fisk, and M.~Z. Hasan:
  Phys. Rev. B {\bfseries 92} (2015) 104420.

\bibitem{Ramankuttya2016}
S.~V. Ramankuttya, N.~de~Jonga, Y.~K. Huanga, B.~Zwartsenberga, F.~Masseeb,
  T.~V. Baya, M.~S. Goldena, and E.~Frantzeskakis: J. Electron Spectrosc.
  Relat. Phenom. {\bfseries 208} (2016) 43.

\bibitem{Koitzsch2016}
A.~Koitzsch, N.~Heming, M.~Knupfer, B.~B$\ddot{\rm u}$chner, P.~Y.
  Portnichenko, A.~V. Dukhnenko, N.~Y. Shitsevalova, V.~B. Filipov, L.~L. Lev,
  V.~N. Strocov, J.~Ollivier, and D.~S. Inosov: Nat. Commun. {\bfseries 7}
  (2016) 10876.

\bibitem{Ohkawa1983}
F.~J. Ohkawa: J. Phys. Soc. Jpn. {\bfseries 52} (1983) 3897.

\bibitem{Ohkawa1985}
F.~J. Ohkawa: J. Phys. Soc. Jpn. {\bfseries 54} (1985) 3909.

\bibitem{Shiina1997}
R.~Shiina, H.~Shiba, and O.~Thalmeier: J. Phys. Soc. Jpn. {\bfseries 66} (1997)
  1741.

\bibitem{Shiina1998}
R.~Shiina, O.~Sakai, H.~Shiba, and O.~Thalmeier: J. Phys. Soc. Jpn. {\bfseries
  67} (1998) 941.

\bibitem{Shiba1999}
H.~Shiba, O.~Sakai, and R.~Shiina: J. Phys. Soc. Jpn. {\bfseries 68} (1999)
  1988.

\bibitem{Hanzawa2000}
K.~Hanzawa: J. Phys. Soc. Jpn. {\bfseries 69} (2000) 510.

\bibitem{Sakurai-Kuramoto2004}
G.~Sakurai and Y.~Kuramoto: J. Phys. Soc. Jpn. {\bfseries 73} (2004) 225.

\bibitem{Sakurai2005}
G.~Sakurai: Dr. Thesis, Tohoku University (2005).

\bibitem{Lu-Huang2017}
H.~Lu and L.~Huang: Phys. Rev. B {\bfseries 95} (2017) 155140.

\bibitem{Barman2019}
C.~K. Barman, P.~Singh, D.~D. Johnson, and A.~Alam: Phys. Rev. Lett. {\bfseries
  112} (2019) 076401.

\bibitem{RK1954}
M.~A. Ruderman and C.~Kittel: Phys. Rev. {\bfseries 96} (1954) 99.

\bibitem{Kasuya1956}
T.~Kasuya: Prog. Theor. Phys. {\bfseries 16} (1956) 45.

\bibitem{Yosida1957}
K.~Yosida: Phys. Rev. {\bfseries 106} (1957) 893.

\bibitem{SL1985}
D.~Schmitt and P.~M. Levy: J. Magn. Magn. Mater. {\bfseries 49} (1985) 15.

\bibitem{Frisken-Miller1986}
S.~J. Frisken and D.~J. Miller: Phys. Rev. Lett. {\bfseries 57} (1986) 2971.

\bibitem{Tanaka2010}
K.~Tanaka, T.~Araki, and K.~Hanzawa: Phys. Rev. B {\bfseries 82} (2015) 134435.

\bibitem{Hanzawa2015}
K.~Hanzawa: J. Phys. Soc. Jpn. {\bfseries 84} (2015) 024717.

\bibitem{w2k2002}
P. Blaha, K. Schwarz, G.K.H. Madsen, D. Kvasnicka, J. Luitz, A Full-Potential
  Linearized Augmented-Plane Wave Package for Calculating Crystal Properties
  (WIEN2k,Vienna University of Technology, 2002).

\bibitem{w2k-Schwarz1990}
K.~Schwarz, P.~Sorantin, and S.~B. Trickey: Comput. Phys. Commun. {\bfseries
  59} (1990) 399.

\bibitem{w2k-Schwarz2002}
K.~Schwarz, P.~Blaha, and G.~K.~H. Madsen: Comput. Phys. Commun. {\bfseries
  147} (2002) 71.

\bibitem{PBE-GGA1996}
J.~P. Perdew, S.~Burke, and M.~Ernzerhof: Phys. Rev. Lett. {\bfseries 77}
  (1996) 3865.

\bibitem{Wyckoff1964}
R.~Wyckoff: {\em Crystal structures} (2nd Ed, John Wiley \& Sons, NewYork,
  1964), pp. 202--203.

\bibitem{MV1997}
N.~Marzari and D.~Vanderbilt: Phys. Rev. B {\bfseries 56} (1997) 12847.

\bibitem{Souza2001}
I.~Souza, N.~Marzari, and D.~Vanderbilt: Phys. Rev. B {\bfseries 65} (2001)
  35109.

\bibitem{Marzari2012}
N.~Marzari, A.~A. Mostofi, J.~R. Yates, I.~Souza, and D.~Vanderbilt: Rev. Mod.
  Phys. {\bfseries 81} (2012) 1419.

\bibitem{w90-Mostofi2008}
A.~Mostofi, J.~R. Yates, Y.-S. Lee, I.~Souza, D.~Vanderbilt, and N.~Marzari:
  Comput. Phys. Commun. {\bfseries 178} (2008) 685.

\bibitem{w2w-Kunes2010}
J.~Kune$\check{\rm s}$, R.~Arita, P.~Wissgott, A.~T.~H. Ikeda, and K.~Helde:
  Comput. Phys. Commun. {\bfseries 181} (2010) 1888.

\bibitem{Yosida-Yamada1986}
K.~Yosida and K.~Yamada: Prog. Theor. Phys. {\bfseries 76} (1986) 621.

\bibitem{Kubo-Kuramoto2003}
K.~Kubo and Y.~Kuramoto: J. Phys. Soc. Jpn. {\bfseries 72} (2003) 1859.

\bibitem{Kubo-Kuramoto2004}
K.~Kubo and Y.~Kuramoto: J. Phys. Soc. Jpn. {\bfseries 73} (2004) 216.

\bibitem{Mannix2005}
D.~Mannix, Y.~Tanaka, D.~Carbone, N.~Bernhoeft, and S.~Kunii: Phys. Rev. Lett.
  {\bfseries 95} (2005) 117206.

\bibitem{Kurasawa2007}
K.~Kuwahara, K.~Iwasa, M.~Kohgi, N.~Aso, M.~Sera, and F.~Iga: J. Phys. Soc.
  Jpn. {\bfseries 76} (2007) 093702.

\bibitem{Matsumura2014}
T.~Matsumura, S.~Michimura, T.~Inami, T.~Otsubo, H.~Tanida, F.~Iga, and
  M.~Sera: Phys. Rev. B {\bfseries 89} (2014) 014422.

\bibitem{Inami2014}
T.~Inami, S.~Michimura, Y.~Hayashi, T.~Matsumura, M.~Sera, and F.~Iga: Phys.
  Rev. B {\bfseries 90} (2014) 041108(R).

\bibitem{Sera2018}
M.~Sera, K.~Kunimori, T.~Matsumura, A.~Kondo, H.~Tanida, H.~Tou, and F.~Iga:
  Phys. Rev. B {\bfseries 97} (2018) 184417.

\bibitem{Ikeda2012}
H.~Ikeda, M.-T. Suzuki, R.~Arita, T.~Takimoto, T.~Shibauchi, and Y.~Matsuda:
  Nat. Phys. {\bfseries 8} (2012) 528.

\bibitem{Nomoto-Ikeda2014}
T.~Nomoto and H.~Ikeda: Phys. Rev. B {\bfseries 90} (2014) 125147.

\bibitem{Nomoto-Ikeda2016}
T.~Nomoto and H.~Ikeda: Phys. Rev. Lett. {\bfseries 117} (2016) 217002.

\bibitem{Georges1996}
A.~Georges, G.~Kotliar, W.~Krauth, and R.~J. Rozenberg: Rev. Mod. Phys.
  {\bfseries 68} (1996) 13.

\bibitem{Rohringer2018}
G.~Rohringer, H.~Hafermann, A.~Toschi, A.~â. Katanin, A.~â. Antipov, M.~â.
  Katsnelson, A.~I. Lichtenstein, A.~N. Rubtsov, and K.~Held: Rev. Mod. Phys.
  {\bfseries 90} (2018) 025003.

\bibitem{Otsuki2015}
J.~Otsuki: Phys. Rev. Lett. {\bfseries 115} (2015) 036404.

\end{thebibliography}


\end{document}